\documentclass[useAMS,usenatbib]{mn2e}
\usepackage{graphicx}
\usepackage{amssymb}\usepackage{hyperref}

\newcommand{\xmm}{{\it XMM-Newton} }

\title[Ripple effect in broad FeK$\alpha$ lines]{Ripple effects \& oscillations in the broad FeK$\alpha$ line as a probe of massive black hole mergers}

\author[B. McKernan, K.E.S. Ford, B. Kocsis \& Z.Haiman]{B. McKernan$^{1,2,3}$\thanks{E-mail:bmckernan at amnh.org (BMcK)}, K.E.S. Ford$^{1,2,3}$, B.Kocsis$^{4,6}$ \& Z.Haiman$^{5}$ \\
$^{1}$Department of Science, Borough of Manhattan Community College, City University of New York, New York, NY 10007, USA\\
$^{2}$Department of Astrophysics, American Museum of Natural History, New York, NY 10024, USA\\
$^{3}$Graduate Center, City University of New York, 365 5th Avenue, New York, NY 10016, USA\\
$^{4}$Harvard-Smithsonian Center for Astrophysics, 60 Garden St., Cambridge, MA 02138, USA\\
$^{5}$Department of Astronomy, Columbia University, New York, NY 10027, USA\\
$^{6}$Einstein Fellow\\
}
\begin{document}

\date{Accepted. Received; in original form}

\pagerange{\pageref{firstpage}--\pageref{lastpage}} \pubyear{2013}

\maketitle

\label{firstpage}

\begin{abstract}
When a sufficiently massive satellite (or secondary) black hole is
embedded in a gas disk around a (primary) supermassive black hole, it
can open an empty gap in the disk. A gap-opening secondary close to
the primary will leave an imprint in the broad component of the
Fe~K$\alpha$ emission line, which varies in a unique and predictable
manner. If the gap persists into the innermost disk, the effect
consists of a pair of dips in the broad line which ripple blue-ward
and red-ward from the line centroid energy respectively, as the gap
moves closer to the primary. This ripple effect could be unambiguously
detectable and allow an electromagnetic monitoring of massive black
hole mergers as they occur. As the mass ratio of the secondary to
primary black hole increases to $q\gtrsim 0.01$, we expect the gap to
widen, possibly clearing a central cavity in the inner disk, which
shows up in the broad Fe~K$\alpha$ line component. If the secondary
stalls at $\geq 10^{2}r_{g}$ in its in-migration, due to low
co-rotating gas mass, a detectable ripple effect occurs in the broad
line component on the disk viscous timescale as the inner disk
drains and the outer disk is dammed. If the secondary maintains an accretion disk within a central cavity, due to dam bursting or leakage, a periodic 'see-saw' oscillation effect is exhibited in the
observed line profile.  Here we demonstrate the range of ripple effect
signatures potentially detectable with \emph{Astro-H} and \emph{IXO/Athena}, and oscillation effects potentially 
detectable with \xmm or \emph{LOFT} for a wide variety of merger and
disk conditions, including gap width (or cavity size), disk
inclination angle and emissivity profile, damming of the accretion
flow by the secondary, and a mini-disk around the satellite black
hole. A systematic study of ripple effects would require a telescope effective area substantially larger than that planned for \emph{IXO/Athena}. Future mission planning should take this into account. Observations of the ripple effect and periodic oscillations can be used to provide an \emph{early warning} of gravitational radiation
emission from the AGN. Once gravitational waves consistent with
massive black hole mergers are detected, an archival search for the
FeK$\alpha$ ripple effect or periodic oscillations will help in
localizing their origin.
\end{abstract}


\section{Introduction}
\label{sec:intro}
Galactic nuclei host supermassive black holes ($>10^{6}M_{\odot}$) \citep{b4} and should 
be the sites of mergers of massive black holes. Major and minor galactic mergers in 
$\Lambda$CDM cosmology should inject supermassive and intermediate mass black holes (BH) into 
the dominant galaxy. Following the merger of two BH-harboring galaxies, the BHs sink to the bottom of the new galactic potential via dynamical friction in approximately a galactic dynamical timescale \citep{bbr80}. In addition to stellar interactions \citep[e.g.][]{preto11}, many studies have shown that gas in the vicinity of the binary could aid in hardening the binary down to $\ll$ pc separations \citep[e.g.][]{esc05,Dotti+07,Mayer+2007,hkm09,Lodato:2009,Cuadra:2009,Nixon2011,CM11}. Black holes of intermediate mass can 
also grow quickly in gas disks in active galactic nuclei (AGN) via stellar/stellar remnant collisions and gas accretion 
\citep{lev07,mck12a} and orbital decay can generate a later merger with the central black hole. Massive black hole mergers should correspond to some of 
the strongest sources of gravitational waves in the Universe. Since we have yet to detect 
gravitational radiation directly, any electromagnetic signature that allows us to track 
massive mergers as they occur will be valuable for the study of strong gravity 
and the growth of the largest black holes in the Universe.

Gravitational torques from satellites in gas disks act to repel gas away from the
 satellite orbit. Sufficiently massive satellites can open empty annular gaps in protoplanetary disks \citep[e.g.][\& references therein]{b2,poll96,ward97,arm10,DM2012}. Analogously, sufficiently massive black 
holes can open gaps in AGN gas disks \citep[e.g.][]{sc95,ivan99,lev07,Dotti+07,hkm09,mck12a,baru12}, leaving a number of unique 
observational signatures \citep{mck12b}. As the ratio of satellite black hole mass to 
central black hole increases to $10^{-2} \leq q \leq 1$, the gap can widen to form a 
central cavity \citep[e.g.][\& references therein]{Artymowicz:1994,liu03,mp05,mm08,dzm12}. A gap-opening black hole in the innermost AGN disk, analogous to a gap-opening 'hot Jupiter' in a protoplanetary disk, can leave an imprint on a relativistically and gravitationally broadened FeK$\alpha$ line profile. The ripple effects and oscillations that result from a variety of black hole merger scenarios: (1) are potentially detectable with near-future X-ray detectors such as \emph{Astro-H},\emph{LOFT} and \emph{IXO/Athena} and possibly with present instruments such as \xmm, (2) can be used to follow massive mergers as they happen, (3) provide advance warning of gravitational radiation from the source \citep{khm08} and (4) can test models of extreme gravity, independent of the detection of gravitational radiation. 
However, detecting the effects outlined here with present and future proposed missions requires serendipity. To systematically survey sites of potentially merging massive binaries covering all the effects discussed here requires at a minimum, the planned effective area of \emph{LOFT} combined with the planned spectral resolution of \emph{IXO/Athena}. 

In section~\S\ref{sec:feka} we review the broad FeK$\alpha$ line and outline the ripple 
effect caused by a narrow annular gap around the secondary's orbit, deep in the potential well of a supermassive 
black hole. We also discuss several other different plausible configurations for the gas distribution around binaries. In section~\S\ref{sec:vars} we demonstrate the effects
 of an annular gap on the broad FeK$\alpha$ line profile for a range of basic observables (angle to observer, X-ray emissivity profile). In section~\S\ref{sec:vars_models} we discuss other possible circumbinary gas disk configurations for merging BHs including stalled migration, damming of the accretion flow, a central cavity and a secondary accretion disk around the gap-opening black hole. In section~\S\ref{sec:gw} we discuss the gravitational wave signals that may correspond to the FeK$\alpha$ broad line profiles. Finally in \S\ref{sec:conclusions} we summarize our main results and present our conclusions.

\section{A ripple effect in the broad FeK$\alpha$ line}
\label{sec:feka}

\subsection{The broad FeK$\alpha$ line component}
\label{sec:broad}
Broad FeK$\alpha$ lines are observed in several nearby AGN 
\citep[e.g.][]{nan97,turner02,reynow03,braito07,mini07}. The narrow core of the 
FeK$\alpha$ line observed in AGN \citep[e.g.][]{yp04} does not correlate significantly with outflowing gas \citep{mck07} and so presumably originates in fluorescing cold  gas \footnote{By 'cold' we mean Fe~{\sc i}-Fe~{\sc xvii}} far from the supermassive black hole \citep{shu10}. We shall not discuss the narrow component further here. However, the broad FeK$\alpha$ line component in (AGN) is widely believed to 
originate in fluorescent Fe deep in the gravitational potential well of the central 
supermassive black hole. The broad FeK$\alpha$ line profile is complicated by 'horns' 
due to relativistic boosting and a strong red wing due to gravitational 
broadening \citep[see e.g.][for a review]{reynow03}. 

There is some recent controversy over the origin of the broad component of the line, with 
some suggestions that it originates in complex absorption \citep[e.g.][]{b58}. However, 
partial covering models do not eliminate the need for a broad FeK$\alpha$ component 
\citep{b15} and Compton-thick outflows are ruled out in the innermost 
regions of the accretion disk \citep{rey12}. Furthermore, the appearance 
of the broad line during transit events in AGN \citep{my98} is consistent with an origin in 
the innermost accretion disk \citep{b14}. So it seems we should expect a broadened Fe~K$\alpha$ line component originating deep in the gravitational potential well around supermassive black holes. 

Although the structure of the innermost accretion flow is unknown, it is common to assume a disk structure. The innermost parts of a disk must lie in the orbital plane of a massive merging binary \citep{ivan99}. In this context, the broad  Fe~K$\alpha$ line profile has been used to test models of warped disks \citep{hb00,fmv05}, thick disks \citep{ww07}, accretion rings \citep{sksd11}, spiral density waves \citep{hb02} and the presence of a binary with two disks \citep{srrd12}.  Given the origin of the broad component of the Fe~K$\alpha$ line in the innermost regions of the AGN disk, a gap or cavity in the inner disk due to the presence of a massive (secondary) black hole will have an effect on the broad line profile. We discuss this effect in detail in the remainder of this paper. 

Following \citet{fab89}, we calculate the flux emitted by a axisymmetric thin disk orbiting a Schwarzschild black hole. The ratio of the emitted energies from a point in the disk to the observed energy is given by \citep{fab89}
\begin{equation}
(1+z)=\left(1-\frac{6M}{r}\right)^{-1/2}\left[ 1 +\frac{\rm{cos} \beta}{[r(1+\rm{tan}^{2}\xi)/M-1]^{1/2}}\right]
\label{eq:1pz}
\end{equation}
where $(r,\phi)$ is the location of the emitting point in the disk (with $r$ in units of $r_{g}=GM/c^{2}$, the gravitational radius). In eqn.~(\ref{eq:1pz}) $M$ is the black hole mass, $\beta$ is the angle between the disk plane and the plane of the photon trajectory, $\xi +\pi/2$ is the angle between the emission of the photon and the line connecting the emitting point to the black hole. The quantity $\rm{cos}\beta$ may be written as \citep{fab89}
\begin{equation}
\rm{cos} \beta=\frac{\rm{cos}\phi \rm{sin}\theta}{[\rm{cos}^{2}\theta +\rm{cos}^{2}\phi \rm{sin}^{2}\theta]^{1/2}}
\label{eq:beta}
\end{equation}
with $\theta$ the angle between the observer and the disk. The specific intensity in the frame of the emitting plasma is approximated as a delta function $I_{\rm{em}}=\epsilon \delta(E-E_{0})$ where $E_{0}$ is the line rest energy. The structure of both the hot corona (source of the X-ray continuum) and the inner disk (source of the fluorescing Fe) is unknown, so $\epsilon$ the disk emissivity is simply assumed to be $\propto (r/M)^{-k}$. Therefore the specific flux measured by the observer from portion $r dr d\phi$ of the disk is then given by \citep{fab89}
\begin{equation}
dF_{\rm {obs}}=\frac{1}{(1+z)^{3}}I_{\rm {em}}\frac{\partial \Omega}{\partial (r,\phi)} r dr d\phi 
\label{eq:flux}
\end{equation}
where the solid angle $d\Omega=D^{-2}\rm{b} db d\phi^{\prime}$ where $D$ is the distance to the disk from the observer, $\rm{b}$ is the photon impact parameter and 
\begin{equation}
\frac{\partial \phi^{\prime}}{\partial \phi}=\frac{\rm{cos} \theta}{\rm{cos}^{2}\phi + \rm{sin}^{2}\phi \rm{cos}^{2}\theta}.
\label{eq:jacobian}
\end{equation}
In the weak field limit, the photon geodesics can be approximated with straight lines, which means that not all relativistic effects are fully calculated at $<20r_{g}$ in this approach \citep{bd04}. By integrating eqn.~(\ref{eq:flux}) as
\begin{equation}
F_{\rm obs}=\int^{2\pi}_{0} \int^{\rm r_{1}}_{\rm r_{inner}} dF_{\rm {obs}} + \int^{2\pi}_{0} \int^{\rm r_{outer}}_{\rm r_{2}} dF_{\rm {obs}} 
\label{eq:integrate}
\end{equation}
where the disk extends from $\rm{r_{inner}}$ to $\rm{r_{outer}}$, with a gap excavated from radii $\rm{r_{1}}$ to $\rm{r_{2}}$, we obtain the observed broad line flux from a disk containing a gap. In the limit as $r_{1} \rightarrow r_{2}$, and if the disk extends to the innermost stable circular orbits (ISCO), $r_{\rm inner} \rightarrow r_{\rm ISCO}$, we recover the observed broad line flux from the total disk. In the limit as $r_{\rm inner} \rightarrow r_{1}$ we recover the observed broad line flux due to the inner disk minus a cavity spanning $r_{\rm ISCO}$ to $r_{2}$.

\subsection{Gap-opening in disks and merging black holes}
\label{sec:gaps}
The disk emissivity which produces the broad Fe~K$\alpha$ line component depends on both the structure of the X-ray continuum emitting corona plus the gas surface density distribution in the circumbinary disk. The geometry of both the corona and inner disk are poorly understood, particularly at the late
stages of a merger with the central supermassive black hole.  The main purpose of this paper is to illustrate that different
possible gas configurations produce very different broad FeK$\alpha$ line
profiles.  Although our illustrations are based on toy models, the
main features should be robust, and potentially distinguishable in observations with {\it
Astro-H} and {\it IXO/Athena}. In this section, we outline possible models for the circumbinary gas, including those with a narrow annular gap, a fully empty central
cavity, or a central cavity with possible minidisks inside the cavity.
Throughout this discussion, we emphasize the very large existing
theoretical uncertainties.

BH binaries are commonly believed to be surrounded by a thin
($H/R\lesssim 1$) circumbinary disk.  In models where the binary
results from a galaxy-galaxy merger, the merger is expected to deliver
the nuclear BHs (e.g. \citealt{springel05,robertson06}), along with
copious amounts of gas \citep{bh92}, to the central regions of the new
galaxy. The gas cools efficiently, and loses angular momentum,
creating a thin sub-parsec disk in which the pair of BHs are embedded
(e.g. \citealt{Dotti+09,hq10,CM11}).
IMBHs with lower masses can also grow quickly in situ in the disk
around a single supermassive black hole, again producing a pair (or more) of BHs embedded
in a thin disk \citep{mck12a}.

A BH binary can exchange angular momentum with the
disk, distorting the disk's density profile, causing the secondary
BH to migrate inward.  In the limit of a very low-mass secondary
($q\equiv M_2/M_1 \ll 1$), the distortions, in the form of spiral
density waves, remain in the linear regime (and the corresponding Type
I migration is very rapid; e.g. \citealt{gt80}).  In this limit, the
FeK$\alpha$ line would be hardly affected by the presence of the
low-mass secondary \footnote{However, if the local gas mass is low enough, or for low enough values of $H/r, \alpha$ in the innermost disk, even low-mass migrators expected in the AGN disk \citep{m11a,mck12a} could open a gap or cavity and affect the Fe~K$\alpha$ line as described here.}. As the mass of the secondary increases, gas is gravitationally torqued away from the orbit, lowering the co-rotating gas density and a particularly rapid form of migration (Type III) may occur \citep{mp03,m11b}. Once the secondary becomes sufficiently massive it will open an empty annular gap in the disk, analogous to gaps known to exist in protoplanetary disks. This happens above the critical mass ratio (e.g. \citealt{arm10}) of
\begin{equation}
q \geq \left( \frac{27 \pi}{8}\right)^{1/2}\left(\frac{H}{r}\right)^{5/2}\alpha^{1/2}
\label{eq:q}
\end{equation}
where $H/r$ is the geometric thickness (aspect ratio) of the disk and
$\alpha$ is the disk viscosity parameter \citep{ss73}. A black hole
with $q \geq 10^{-4}$ should be able to open a gap in a fiducial AGN disk.
This mass range spans massive stars or black holes of $\geq 10^{2}M_{\odot}$ in accretion disks around $\sim 10^{6}M_{\odot}$ supermassive black holes
\citep{mck12a}, all the way up to supermassive black hole binaries
that form long after a major galactic merger has occurred.  The radial extent of
the gap would be of order a few Hill radii ($\pm R_H$).  In this case,
the secondary can be regarded as a particle in the disk, migrating
inward on the viscous time-scale (so-called Type II migration).

When the secondary arrives at the radius where the local disk mass is
too small to absorb the secondary's angular momentum, its migration
must stall (or at least slow down).  An important point is that
``standard'' thin disks have relatively low mass, and this stalling
occurs at fairly large binary separations. With the exception of very
unequal masses $q\lesssim 0.01$, the transition takes place well in
the outer region of the disk, at $r>$several $\times 10^3 r_g$.  Even
for $q\ll 1$ binaries, the transition is located at
\begin{equation}
r_{\rm stall}=260 (M_1/10^7{\rm M_\odot})^{-2/7}(q/10^{-4})^{2/7}
\end{equation}
Schwarzschild radii, which may lie outside the region of interest for producing broad FeK$\alpha$ lines depending on $r_{\rm outer}$ (the
above equation assumes standard disk parameters; see eqn. 25 in
\citealt{hkm09} for the full expression). Once inside this stalling 
radius, the secondary is no longer able to maintain steady Type II
migration on the viscous time-scale, and the evolution of the
binary+disk system from this point onward becomes very poorly known.
The inner disk - whose viscous time is short - is believed to drain
onto the primary, creating a central cavity. Continued accretion from
larger radii causes a pile-up of gas just outside the secondary's
orbit, analogous to a dam in a river
\citep{sc95,ivan99,mp05,Chang+2010,Rafikov2012}.
If the dam is 100\% efficient in continuing to block the gas arriving
from large radii, then the binary, at small separations, will be
surrounded by an empty inner cavity, devoid of any gas.  Such inner
cavities have been seen in many numerical simulations (starting
with the seminal work by
\citet{Artymowicz:1994}).  However, simulations do not follow the system
on long enough timescales to determine whether such a cavity could
persist for many viscous times. In the face of ongoing accretion, an
initial cavity may be gradually refilled (if the ``dam'' is porous),
or else the gas accumulated outside the orbit may eventually cause the
dam to ``burst'' \citep{Kocsis+2012a,Kocsis+2012b}.

An empty cavity would of course strongly modify the shape of the
FeK$\alpha$ line.  However, this picture is oversimplified.  Many
numerical simulations imply that gas can enter such a cavity
in narrow collimated streams
(\citealt{Artymowicz:1996,Hayasaki:2007,mm08,Cuadra:2009,roedig11,Roedig:2012:arXiv,ShiKrolik:2011}).
Such streams may feed a ``minidisk'' around the secondary and perhaps
also the primary BH \citep{hay08,srrd12,FarrisGold:2012}. A minidisk around the secondary may be favored, but the details depend
on the angular momentum and shocks in the material in the streams.
The streamers and minidisks could create some additional FeK$\alpha$
emission. For example, if both disks emit, there would be a double
FeK$\alpha$ line (\cite{srrd12}).  On the other hand, it is unclear
whether such streams and minidisks may persist down to small binary
separations.  At separations of $\lesssim 200 r_g q^{2}$ (see eqn. 30 in
\citealt{hkm09} for the full expression, including dependencies on disk parameters), 
the binary BH emits GWs efficiently and is driven to merger by this GW
emission. It effectively decouples from the disk (\cite{liu03,mp05})
and ``runs away''.  The merger timescale  is given by \citep{pet64}
\begin{equation}
\tau_{\rm GW} \approx 10^{12} \rm{yr} \left(\frac{10^{3}M_{\odot}}{M_{2}}\right)\left(\frac{10^{6}M_{\odot}}{M_{1}}\right)^{2} \left(\frac{a}{0.001 \rm{pc}}\right)^{4} (1-e^{2})^{7/2}
\label{eq:gravrad}
\end{equation}
where $M_{2}$ is the mass of the secondary, $M_{1}$ is the mass of the primary, $a$ is the binary separation and $e$ is the secondary orbital eccentricity. For a secondary on a circularized orbit 
with $q=3 \times 10^{-3}$ located at $a=(10)100r_{g}$ in an AGN disk,
$\tau_{\rm GW} \sim (30\rm{yr})0.3\rm{Myr}$ around a $10^{8}M_{\odot}$ SMBH and 
$\tau_{\rm GW} \sim (1\rm{yr})0.01\rm{Myr}$ around a $10^{6}M_{\odot}$ SMBH respectively. Simulations in the relativistic regime found that
gas streams can follow the binary to small radii ($\sim$ few $R_s$;
e.g. \citealt{Noble+2012,FarrisShap:2011,FarrisGold:2012}), but these
simulations start with small binary separations, with gas assumed to
have followed the BH to these initial separations. 

As emphasized recently by \citet{Kocsis+2012a,Kocsis+2012b}, even
apart from possible streamers and minidisks inside the central cavity,
the formation of the cavity itself, at large radii, is still
uncertain. The coupled time-dependent processes of the secondary's
migration, inner cavity formation, and damming-up of gas outside the
secondary's orbit, has not been modeled self-consistently, even in
one-dimensional calculations.  However, using self-consistent {\em
steady-state} solutions,
\citet{Kocsis+2012b} have argued that in many cases, pile-up can cause
a dam overflow at large binary separations.  In this case, when
the binary arrives at small radii relevant to the FeK$\alpha$ line,
the gas configuration would presumably resemble a filled disk, with an
annular gap (of order the Hill radius).  Although the above is yet to
be demonstrated by a self-consistent calculation, it is at least
suggested by recent simulations of \cite{baru12}.  This is the only
existing simulation that looks at small binary separations, in the
GW-driven regime, and does not manually insert an empty cavity in the
beginning.  They show that the secondary moves inward, ``ice-skating''
across the inner disk (with the inner disk material continually
crossing the secondary's orbit outward, on horse-shoe orbits).  This
simulation is in 2D, and assumes a constant kinematic viscosity physics, which may not be accurate. Nevertheless, it suggests that if the cavity filled
in at large radius, it will remain filled until the end, except for a
narrow annular gap around the secondary.

We are motivated by the above discussion to consider the impact on the Fe~K$\alpha$ line of four different circumbinary gas disk geometries, all variations on a standard single-BH accretion disk: (i) a narrow annular gap around the secondary, (ii) an empty central cavity, (iii) a central cavity surrounded by an overdense ring ('dam'), and (iv) a central cavity with a small, circumsecondary accretion disk ('minidisk') inside it. We regard (i) as our fiducial case, addressing it first in \S\ref{sec:intro_ripple} and \S\ref{sec:vars}. We then examine the alternative scenarios (ii)-(iv) in \S\ref{sec:vars_models}.

\subsection{A ripple effect in the broad FeK$\alpha$ line component}
\label{sec:intro_ripple}

\begin{figure}
\includegraphics[width=3.35in,height=3.35in,angle=-90]{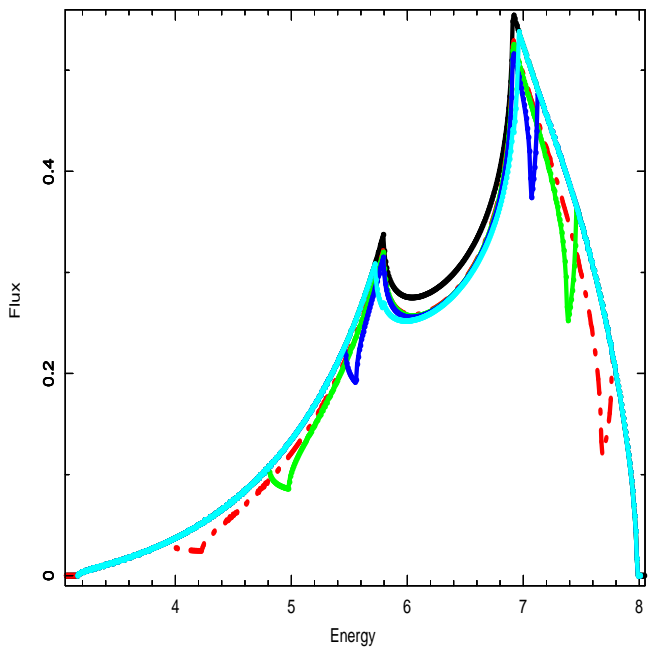}
\caption{From \citet{mck12b}, showing Iron line flux in arbitrary units versus Energy in 
units of keV,binned at approximately the energy resolution ($\sim 7$eV) expected for \emph{Astro-H}. The curves demonstrate the progression of the 'ripple effect' in the 
Fe~K$\alpha$ line profile given the in-migration of a gap-opening black hole (mass ratio 
$q=3 \times 10^{-3}$) located at R in the inner disk. The empty gap has width $2 R_{H}$ 
where $R_{H}=(q/3)^{1/3}R$. The disk has an assumed X-ray emissivity radial profile of 
$r^{-2.5}$ and the angle to the observer's sightline is $\theta=60^{o}$ (where 
$\theta=0^{o}$ is face-on). The solid black 
curve indicates the unperturbed Fe~K$\alpha$ line profile (i.e. no gap). The turquoise 
solid curve corresponds to the same Fe~K$\alpha$ line but with an annulus at $90\pm9r_{g}$ in the inner disk. Dark blue solid curve denotes an annulus at $50\pm 5 r_{g}$, Green 
solid curve 
corresponds to an annulus at $20\pm 2r_{g}$ and the dashed red curve corresponds to 
an annulus at $10\pm 1r_{g}$.
\label{fig:feka}}
\end{figure}
As a gap-opening black hole migrates inward in an AGN disk, the empty gap spans parts of the disk that produce ever bluer and redder parts of the broad Fe~K$\alpha$ line. Therefore we expect flux 'notches' to be removed from the broad line profile. As the annulus moves inward on the disk viscous timescale, we expect the notches to ripple red-ward and blue-ward respectively from the line centroid. Fig.~\ref{fig:feka} (from \citet{mck12b}) demonstrates this ripple effect for a gap-opening black hole located at circularized orbit radius $R$ centered in 
an empty gap of width $2R_{H}$ where $R_{H}=(q/3)^{1/3}R$. The black hole migrates 
inward with the gap from $100r_{g}$ to $6r_{g}$. We assumed $q=3 \times 10^{-3}$ for ease of calculation, the disk is oriented
 at $60^{o}$ to the observer's line of sight, the X-ray emissivity profile goes as $r^{-2.5}$ and the broad line flux is assumed to come from $<100r_{g}$ in the disk (i.e. $r_{\rm inner}=6r_{g},r_{\rm outer}=100r_{g}$). If we change $r_{\rm outer}$ to larger radii \citep[e.g.][]{NiedZyc2008,NiedMik2010}, the broad line profiles depicted throughout this paper would appear more centrally peaked with 'horns' closer together, resulting in a slight decrease in the prominence of the notches and oscillations described here. If $r_{\rm outer}$ is larger, we become sensitive to stalling radii, gaps and cavities further out in the accretion disk. Exactly how sensitive to detection of the effects discussed here depends on the line emissivity ($\propto r^{-k}$) profile. If the emissivity profile is steep ($k>3$) then most line flux originates close to the BH and we have low sensitivity to features in the disk much further than $\sim 100r_{g}$.  If the emissivity profile is shallower ($k \sim 2$), for large $r_{\rm outer}$, the line 'horns' correspond to material around $\sim 100-150r_{g}$, so gaps at these radii could be detected. In general, our simple assumption that the broad line flux comes from $\lesssim 10^{2}r_{g}$), is reasonably consistent with fits to many observed broad line components  \citep{reynow03}. Likewise, in the following discussion we do not correctly treat general relativistic effects such as light-bending, blurring and BH spin \citep[e.g.][]{Dov04,Bren09,Daus10}. These effects, if properly treated, will tend to smear out sharper notches in the red wing of the broad FeK$\alpha$ line and modify line flux originating from $\lesssim 20r_{g}$. Even though we ignore the many possible confounding effects (e.g. continuum absorption, the location and variation of the continuum source, broken axi-symmetry), we expect that the kinds of spectral features here are broadly robust and allow us to take a first step in characterizing different scenarios of merging massive black holes in the inner accretion disk.

From Fig.~\ref{fig:feka}, the progression of the ripple effect is quite clear. As the satellite black hole and gap migrate into the 
inner $100r_{g}$ of the disk, a flux deficit appears (turquoise solid curve) compared to 
the unperturbed profile (black solid curve). As migration continues, matching notches are 
removed from the broad red wing and the blue wing of the profile, with the deepest notches
 in the blue wing. The twin notches move apart, one red-ward, the other blue-ward further 
from the line centroid (6.40keV) in a distinctive monotonic progression. As the gap moves 
inward in the disk from 50$r_{g}$ (dark blue curve) to 20$r_{g}$ (green curve) to 
10$r_{g}$ (red curve), the blue notch moves from 7.1keV to 7.6keV in the blue wing, and 
the red notch moves from $5.6$keV to 4.2keV in the broad red wing. For a $3000M_{\odot}$ gap-opening IMBH around a $10^{6}M_{\odot}$ supermassive black hole, the progression from green to red curve in 
Fig.~\ref{fig:feka} via GW emission takes (eqn.~\ref{eq:gravrad}) $\sim 16$yrs and from red curve to final merger takes $\sim 1$yr. Because of the shorter timescales, low mass AGN exhibiting broad Fe~K$\alpha$ 
line profiles (e.g. MCG-6-30-15) are the prime targets for observations searching for this ripple effect, and the massive mergers that cause it.

AGN that display this ripple effect in the broad Fe~K$\alpha$ line will be some of the most luminous sources of gravitational radiation 
in the Universe. By searching for the latter stages of a progression as in 
Fig.~\ref{fig:feka}, we can follow the final stages of massive mergers 
\emph{electromagnetically}. The final stages of a ripple effect in the broad FeK$\alpha$ line will provide advance warning of an outburst of gravitational waves from this AGN. 
Conversely, once gravitational waves consistent with an IMBH-supermassive black hole 
merger are actually detected, an archival search for the FeK$\alpha$ ripple effect can be 
used to localize the source. 

We expect departures 
from the predicted progression in Fig.~\ref{fig:feka} at very small radii either as the 
disk profile changes or as the gravitational 
radiation luminosity increases dramatically and perturbs the disk ($\alpha$ and $H/r$ 
may change considerably). So the behaviour of the broad component of the 
Fe~K$\alpha$ line may be complicated in the final stages of the merger (in the 
limit of small disk radii). Variations from the predicted profile at small radii will 
include the effects of gravitational radiation on the accretion disk. So tracking the 
ripple effect may allow us to \emph{directly} test models of massive mergers and 
gravitational 
radiation, without actually detecting gravitational radiation. The Fe~K$\alpha$ band 
can also be spectrally complicated by emission lines. By monitoring the 
monotonic change in energy of the blue and red dips, we can break the degeneracy 
associated with high energy Fe emission lines superimposed on the FeK$\alpha$ complex \citep[e.g.][]{my04}. 
Curious 'notch'-like features have been observed in broad Fe~K$\alpha$ line components \citep[e.g.][]{ys05}, but the collecting area and the energy resolution of the high energy 
transmission gratings on board \emph{Chandra} are insufficient to resolve these features and 
follow them over time. The broad component of the Fe~K$\alpha$ line can vary quite significantly over time \citep[e.g.][]{iwa96,dem09,sb12}, but it has been difficult to determine the nature of the variability given detector limitations. 

The energy resolution of the micro calorimeter planned for the \emph{Astro-H} mission will be $\sim 7$eV in the FeK$\alpha$ band \citep{b29}. This exceptional resolution combined with moderate collecting area should be sufficient to distinguish between absorption features and narrow, monotonically migrating notches of the kind depicted in Fig.~\ref{fig:feka} in long, repeated observations of nearby, X-ray bright AGN \citep[e.g.][]{nan97}. However, it is important to note that Fig.~\ref{fig:feka} displays only the predicted broad line flux. To systematically obtain broad line spectra that can distinguish between the different cases in Fig.~\ref{fig:feka} requires a minimum telescope effective area of that planned for \emph{LOFT} ($\sim 10\rm{m}^{2}$ at 6.4keV), \emph{together} with the planned spectral resolution of ($\sim 5-7$eV at 6.4keV) for \emph{Astro-H} and \emph{IXO/Athena}. \citet{vf04} obtain $\sim 18,000$ counts in the broad red wing of the Fe~K$\alpha$ line in a $\sim 320$ks observation of MCG-6-30-15 with the EPIC detector on board \xmm, yielding $\sim 600$ average counts at the limiting energy resolution of $\sim 0.1$keV per bin and very high signal to noise. An observation with a future Large Area High Resolution (L.A.H.R.) telescope with the effective area of \emph{LOFT} would get $\sim 6,000$ average counts per $\sim 0.1$keV bin at very high signal to noise. In this case, observations with \emph{LAHR} would have the signal-to-noise of \xmm EPIC at binnings of $\sim 10$eV, which would be sufficient to at least systematically track the monotonic changes in the energy of the notches in Fig.~\ref{fig:feka}.  In the remainder of this paper we plot all figures at the $\sim 7$eV energy resolution expected for \emph{Astro-H} (and \emph{IXO/Athena}), to illustrate the details that might be observed in the Fe~K broad line complex at sufficiently high signal-to-noise with future detectors.  As we shall show below, the EPIC-pn detector on board \xmm is most useful in testing models of oscillations in the broad Fe~K$\alpha$ line due to secondary mini-disks, until the advent of \emph{LOFT}. In the case of oscillations, high energy resolution is less important than binned-up count-rates and the superior effective area of EPIC makes this the present detector of choice and we recommend that observers search for serendipitous oscillations in the Fe~K band when observing with this instrument. The planned \emph{IXO/Athena} mission will have both high energy resolution \emph{and} large effective area. However, systematic testing of models of massive black hole mergers via the broad Fe~K$\alpha$ line will require Large Area High Resolution (\emph{LAHR}) telescopes with minimum effective area that proposed for \emph{LOFT}, together with the spectral resolution planned for \emph{IXO/Athena}. We recommend that future mission planning take this into consideration. 
 
A related issue to detectability is the expected rate of occurence of binaries (of mass ratio $q$) in AGN. There are at present a small number of bright, nearby AGN that exhibit broad Fe K$\alpha$ lines. Even with an order of magnitude increase in telescope effective area, our sample size would be $\sim 10^{2}$ objects. The probability of seeing a close binary  per AGN is $N_{2}t_{\rm res}N_{\rm obs}/t_{\rm disk}$ where $N_{2}$ is the number of secondaries (for a given range of mass ratios) in the AGN disk over its lifetime ($t_{\rm disk}$),  $t_{\rm res}$ is the residence time of the secondary at the range of disk radii where we can observe the effects discussed here and $N_{\rm obs}$ is the number of minimal S/N exposures of the AGN of duration ($t_{\rm min}$) in which the effect could be detected. For the lowest mass (Type I) migrators we might expect $N_{2} \sim 10^{4}$ in an AGN disk that lasts for $t_{\rm disk} \sim $10Myrs around a $10^{8}M_{\odot}$ SMBH, where $t_{\rm res} \sim 1$yr \citep{mck12a}. For a fiducial \emph{LAHR} telescope, we might expect $t_{\rm min} \sim 10$ks so in a total exposure over a mission lifetime of 1Ms, we would have a $10\%$ chance of detecting a low-mass migrator in any one AGN. For SMBH binaries, $t_{\rm res} \sim 10^{6}$yr or more, so as long as $N_{2}$ for massive SMBH is not vanishingly small, we will also have a reasonable likelihood of detecting SMBH binaries with future missions. For SMBH-IMBH binaries, $N_{2}$ is likely to be of order several \citep{mck12a}, $t_{\rm res}$ is likely to be $\sim 10^{4}$yr in the example above, so in a large sample of X-ray bright AGN, we may  detect a small number of these binaries via effects in the broad Fe~K$\alpha$ line. 

\section{Broad FeK$\alpha$ line profiles: basic observables}
\label{sec:vars}
In this section we discuss the effects of a few basic observables on the expected broad Fe~K$\alpha$ line component profile, including angle to the observers' sightline($\theta$) , assumed X-ray emissivity profile ($\propto r^{-k}$) and the width ($\Delta r=r_{2}-r_{1}$) of the annulus gap. In practice, the broad profiles discussed
 here will be those residuals in the Fe~K band after the subtraction of a narrow (very slowly varying) Fe~K$\alpha$ component due to cold fluorescing material far from the supermassive black hole \citep[e.g.][]{yp04}. The Fe~K line complex actually includes two components (K$\alpha_{1}$ and K$\alpha_{2}$ separated by $13$eV), centered at 6.40keV and an additional Fe~K$\beta$ component at 7.06keV. The Fe~K$\beta$ line should have $\sim 13\%$ of the normalized 
flux of the FeK$\alpha$ line \citep{palm03} and so adds a small percentage change in the profile of the 
overall broad Fe~K line, particularly on the blue-wing. However, the line profiles that we generate here, distinguishing between various merger situations, are not materially effected by the addition of a broad Fe~K$\beta$ component, so we shall not discuss that component further here. Furthermore, although individual highly ionized line components can appear in Fe~K band spectra, these are sufficiently narrow that we can hope to disentangle their signature from broad components with \emph{Astro-H}. For the purposes of the present discussion, in order to understand the sense in which different binary disk geometries affect the broad Fe~K line component, we shall ignore the contribution of highly ionized Fe to the broad Fe~K$\alpha$ line complex.

We calculate the broad FeK$\alpha$ line profile around a Schwarzschild black hole using a modified version of the algorithm that 
generates the \textsc{XSPEC} \textsc{diskline} model \citep{fab89}, described in \S\ref{sec:broad} above. We modify the 
\textsc{diskline} algorithm by subtracting annuli of particular widths ($\Delta r= r_{2}-r_{1}$) or by 
introducing additional emission due to damming of the accretion flow or accretion disks around the secondary black hole (see discussion below). For ease of comparison of basic observables, in this section we discuss an empty gap opened at 
$20\pm 2r_{g}$ in an accretion disk around a supermassive black hole (the green solid 
curve in Fig.~\ref{fig:feka}), where all the broad line emission is assumed to come from disk radii spanning [$r_{\rm inner}=6r_{g},r_{\rm outer}=100r_{g}$]. We then change basic parameters in order to understand 
the sense in which the ripple effect changes. In section~\S\ref{sec:angle} we consider the impact of disk inclination angle ($\theta$) on the observed ripple effect. In section~\S\ref{sec:q} we discuss how changing X-ray emissivity ($\propto r^{-k}$) as a function of radius as well as empty gaps of different widths ($\Delta r$) alters the ripple effect. In section~\S\ref{sec:spin} we briefly discuss how the spin of the central black hole might change observations of the ripple effect. 

\subsection{Changing inclination angle}
\label{sec:angle}
There is a relatively wide range of viewing angles ($\theta$) possible for observations of Seyfert 1 AGN nuclei 
that display broad FeK$\alpha$ lines \citep{nan97}. Indeed if disks are warped close to a spinning black hole, even face-on AGN may contain highly inclined inner disks \citep[e.g.][]{bp75,kp06,nix12}. The most famous example of a broad 
Fe~K$\alpha$ line in an AGN (MCG-6-30-15) is best-fit with disk-inclination angle of 
$\theta=27^{o}$ \citep{reynow03}, where $\theta=0^{o}$ corresponds to face-on.

\begin{figure}
\includegraphics[width=3.35in,height=3.35in,angle=-90]{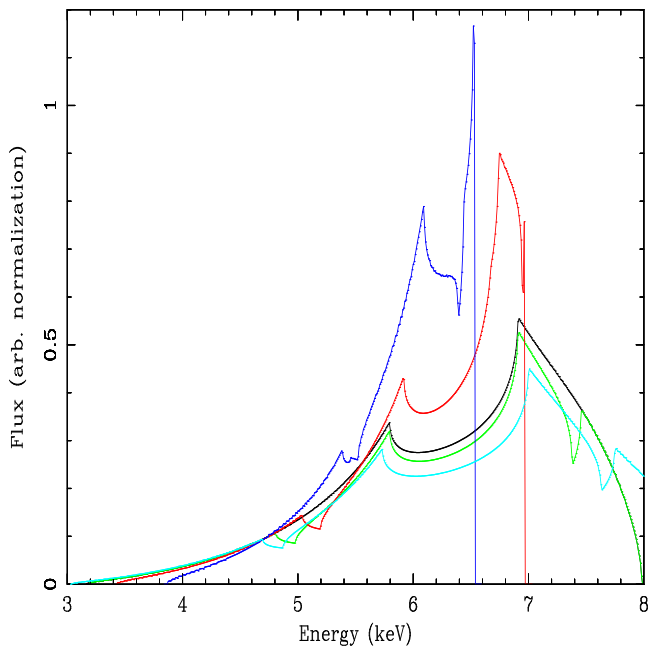}
\caption{Taking the 'ripple' due to an annulus at $20 \pm 2r_{g}$ (the green solid curve 
in Fig.~\ref{fig:feka}) and changing the viewing angle ($\theta$) to the observer. The 
black and green curves correspond to the black and green curves in Fig.~\ref{fig:feka} 
respectively ($\theta=60^{o}$, no gap \& gap). The dark blue, red and light blue curves correspond to the green curve observed at $\theta=20,40,80^{o}$ respectively. All curves are binned at approximately the energy resolution ($\sim 7$eV) expected for \emph{Astro-H}.
\label{fig:varytheta}}
\end{figure}

Fig.~\ref{fig:varytheta} shows the effect of varying inclination angle ($\theta$) on the 
ripple due to the annulus at $20 \pm 2r_{g}$ (green curve) in Fig.~\ref{fig:feka}. From 
Fig.~\ref{fig:varytheta}, as the angle to the observers' sightline is 
increased to $\sim 80^{o}$ (light blue curve), the FeK$\alpha$ line width increases and 
the 'notches' become more prominent. The 'blue' notch shifts away from the H-like 
Fe~{\sc{xxvi}} Ly$\alpha$ line at 6.97keV which prevents possible line confusion. However, 
at this inclination angle, we should expect that intervening material (the dusty torus, 
an outflowing wind or obscuring clouds) should absorb softer X-rays, influencing the shape
 of the continuum and therefore the broad component of the Fe~K$\alpha$ line. Thus, 
unless the AGN is 'naked' (i.e. without obscuring torus or clouds), or the inner disk is 
highly warped, a broad FeK$\alpha$ line at this inclination should be difficult to 
disentangle from absorption effects. 

From Fig.~\ref{fig:varytheta} as $\theta$
decreases to nearer face-on (dark blue and red curves), the likelihood of obscuring 
structures decreases, but the blue wing of the Fe~K$\alpha$ shrinks back towards the 
rest-frame line energy (6.40keV). As a 
result the 'blue' notch becomes much narrower (and harder to detect with \emph{Astro-H}). 
The 'red' notch remains prominent and should stand out clearly in observations with 
\emph{Astro-H}, until obscuring material starts to absorb the continuum significantly at 
this energy. 

\subsection{Changing X-ray emissivity \& gap width}
\label{sec:q}
The geometric structure of the corona (source of the X-ray continuum) and that of the inner disk (source of the fluorescent Fe) are unknown. Therefore the X-ray emission from the inner disk is simply parameterized as a radial powerlaw form ($r^{-k}$), 
where likely ranges for the emissivity span $k \sim 1.5-3$ \citep[e.g.][]{reynow03}. The 
most famous example of a broad Fe~K$\alpha$ line is best-fit with X-ray emissivity profile of $k \sim 3$ \citep{reynow03}.

\begin{figure}
\includegraphics[width=3.35in,height=3.35in,angle=-90]{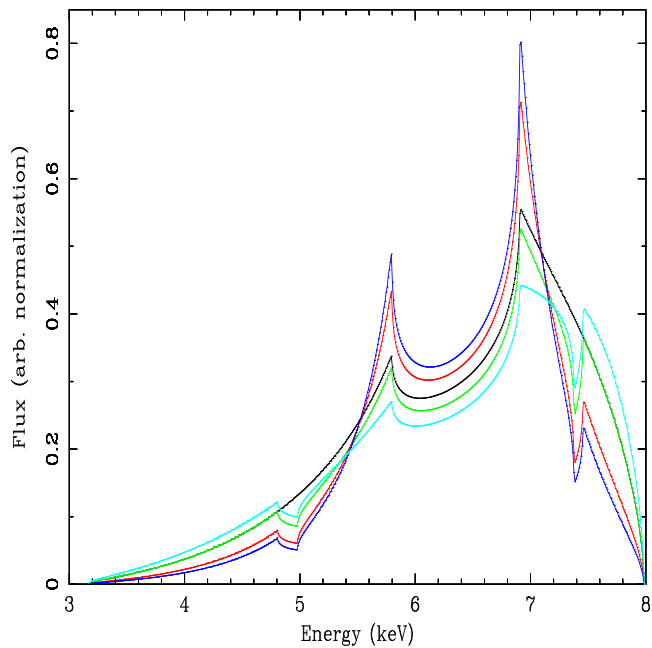}
\caption{Taking the 'ripple' due to an annulus at $20 \pm 2r_{g}$ (the green solid curve 
in Fig.~\ref{fig:feka}) and changing the X-ray emissivity profile ($r^{-k}$). The black 
and green curves correspond to the black and green curves in Fig.~\ref{fig:feka} 
respectively (with $k=2.5$). The dark blue, red and light blue curves correspond to 
$k=1.8,2.0,2.8$ respectively. All curves are binned at approximately the energy resolution ($\sim 7$eV) expected for \emph{Astro-H}.
\label{fig:varyk}}
\end{figure}

Fig.~\ref{fig:varyk} shows the effect of varying the index (k) of the X-ray emissivity powerlaw ($(r/M)^{-k}$), on the ripple due to the annulus at $20 \pm 2r_{g}$ (green curve) in Fig.~\ref{fig:feka}. From 
Fig.~\ref{fig:varyk}, as $k$ gets larger, more and more line flux comes from the innermost regions of the disk, the larger the notches in the line become, and the more pronounced the ripple effect. If $k>3$ due to the special case of magnetic torquing \citep{reynow03}, the late-stage ripple effect will become even more apparent in observations.

\begin{figure}
\includegraphics[width=3.35in,height=3.35in,angle=-90]{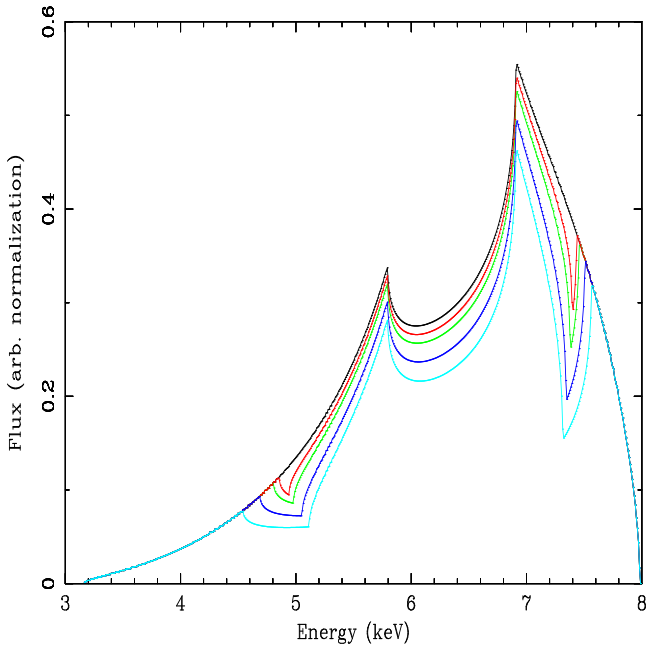}
\caption{Taking the 'ripple' due to an annulus at $20 \pm 2r_{g}$ (the green solid curve in Fig.~\ref{fig:feka}) and changing the annulus width. The black and green curves correspond to the black and green curves in Fig.~\ref{fig:feka} respectively (with annulus located at $20\pm2r_{g}$). The red, dark blue and light blue curves correspond to annuli located at $20\pm 1r_{g},\pm 4r_{g},\pm 6r_{g}$ respectively. All curves are binned at approximately the energy resolution ($\sim 7$eV) expected for \emph{Astro-H}.
\label{fig:varyann}}
\end{figure}

Fig.~\ref{fig:varyann} shows the effect of varying gap width on the 
ripple due to the annulus at $20 \pm 2r_{g}$ (green curve) in Fig.~\ref{fig:feka}. From 
Fig.~\ref{fig:varyann}, as we should expect, the notches in the Fe~K$\alpha$ line profile 
become more pronounced (detectable) as the gap width increases. In \S\ref{sec:vars_models} below we discuss in more detail the effect on the broad Fe~$\alpha$  line profile of increasing the gap width via disk drainage to form a cavity, damming the inflowing gas, and the possibility of an accretion disk around the secondary black hole.

\subsection{Changing black hole spin}
\label{sec:spin}
In the discussion above, we have discussed broad Fe~K$\alpha$ line profiles that originate 
in fluorescent Iron deep in the potential well of a non-spinning ($a=0$, Schwarzschild) 
black hole. In this case, the accretion disk extends inward as far as the innermost 
stable circular orbits (ISCO), which for $a=0$ corresponds to $r_{\rm ISCO}=6r_{g}$ 
\citep{bpt72}. Maximally prograde spinning black holes ($a \sim M$) 
allow the inner edge of the accretion disk to extend practically to the event horizon 
$r_{\rm ISCO} \sim r_{g}$. By contrast, black holes with retrograde spin compared to the disk ($a \sim -M$) have $r_{\rm ISCO} \sim 9r_{g}$. For all except the final stages of merger, the 
spin will make little difference to the ripple effect. However, around a maximally 
spinning black hole, the Fe~K$\alpha$ line will extend redward to very low energies 
($\leq 3$keV,\citet{iwa96}). In the final stages of merger, a gap-opening black hole 
could leave an imprint (at $<6r_{g}$) on a very broad Fe~K$\alpha$ line. Around a low 
mass, rapidly spinning black hole such as in MCG-6-30-15 
($\sim 10^{6}M_{\odot}$, \citet{mck10}), the very late stages of a ripple effect could 
be observed, but would last only a few weeks. A gravitational wave trigger may therefore be useful in finding these late stage events. The possibility is intriguing, since 
tracking a massive merger electromagnetically to the very final stages would be a 
profoundly important advance in this field.

\section{Testing merger models with the broad FeK$\alpha$ line profile}
\label{sec:vars_models}
In this section we examine alternative models for the geometry of the innermost AGN disk, as discussed in \S~\ref{sec:gaps} above, and ask how we might distinguish between these different models using the broad Fe~K$\alpha$ line component. In \S\ref{sec:cavities} we discuss cavity formation and how we might test the formation and presence of cavities using the broad FeK$\alpha$ line profile. In \S\ref{sec:dam} we discuss observational consequences of 'damming' the disk due to pile-up of inflowing gas behind a stalled, migrating secondary black hole. In \S\ref{sec:minidisk} we discuss the possibility of a disk around the secondary black hole and the effect this has on the overall broad Fe~K$\alpha$ line profile. In reality, a combination of all three effects is likely to be present simultaneously.

\subsection{Detecting cavities in the inner AGN disk}
\label{sec:cavities}
\begin{figure}
\includegraphics[width=3.35in,height=3.35in,angle=-90]{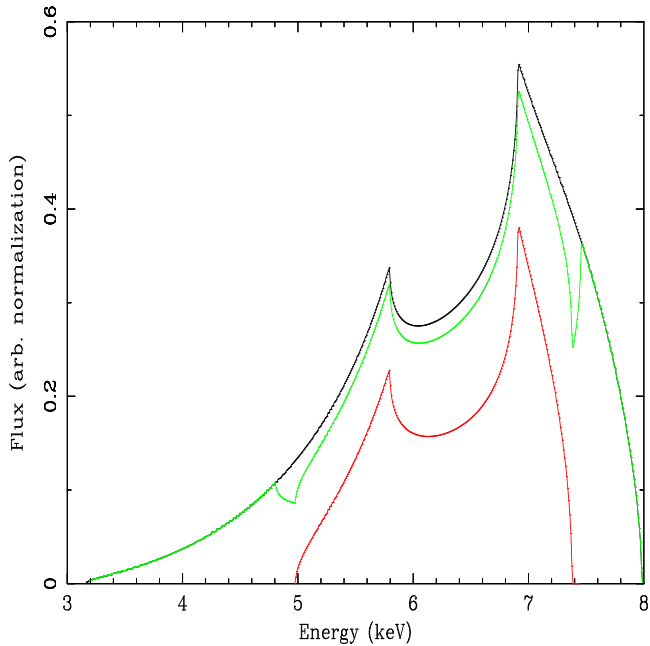}
\caption{Comparing the 'ripple' profile due to an annulus at $20 \pm 2r_{g}$ (the green 
solid curve in Fig.~\ref{fig:feka}) with the line profile due to a disk minus an empty cavity at $r<22r_{g}$. The black and green solid curves correspond to the black and green solid curves in Fig.~\ref{fig:feka} respectively. The red curve corresponds to the FeK$\alpha$ profile from a disk with the same emissivity ($r^{-k}$) and inclination angle ($\theta$), but with an empty cavity at radii $<22r_{g}$. All curves are binned at approximately the energy resolution ($\sim 7$eV) expected for \emph{Astro-H}.
\label{fig:cavity}}
\end{figure}

Fig.~\ref{fig:cavity} shows the effect of a cavity in the innermost disk on the 
ripple due to the annulus at $20 \pm 2r_{g}$ (green curve) in Fig.~\ref{fig:feka}. In Fig.~\ref{fig:cavity} we demonstrate the effect of removing (keeping) all the fluorescing Fe between $6-18r_{g}$ in the red curve (green curve). Note that while this situation corresponds to a stalled migrator at $20r_{g}$, GW emission is likely to rapidly re-start the stalled binary. We choose this value of the separation to simply demonstrate differences in line profiles. The effects are apparent on the broad Fe~K$\alpha$ line profile, where the red and blue wings in the green curve, outside the notches (at $\sim 5.0,7.4$keV respectively), vanish from the red curve. One problem that becomes 
apparent from Fig.~\ref{fig:cavity} is that it will be more difficult to detect the effect 
on the Fe~K$\alpha$ line of black holes that create inner cavities than gap-opening black 
holes. This is not only because the obvious features (the notches) are gone, but also 
because spectral model fitting involves independent flux normalizations for both the 
continuum and the Fe~K$\alpha$ line. Thus, an 
observation of a line profile best fit by the red curve in 
Fig.~\ref{fig:cavity}, could be produced by a satellite black hole in a cavity, but could 
also be produced by a complete disk observed at a different angle ($\theta$) or with a 
shallower emissivity profile (e.g. compare with the dark-blue curve in Fig.~\ref{fig:varyk} corresponding to $r^{-1.8}$). In this case, model degeneracy can be broken by tracking the width of the line profile over time. If a line profile similar to the red curve in Fig.~\ref{fig:cavity} was observed in an AGN, a cavity should decrease in radius and therefore the fluorescent line width should increase monotonically on the disk viscoust imescale. Measurements from other broad lines in the same AGN may also constrain the angle ($\theta$) to the observer of the innermost disk and rule out e.g. near face-on disks.

\begin{figure}
\includegraphics[width=3.35in,height=3.35in,angle=-90]{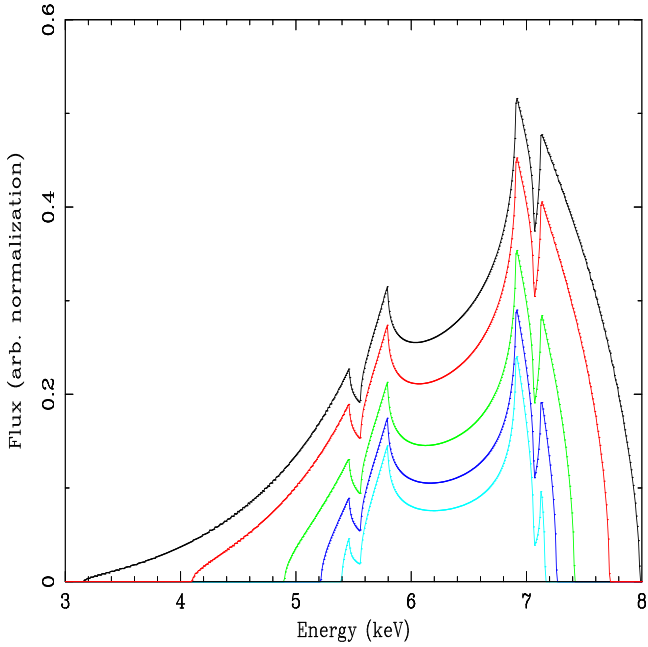}
\caption{Comparing the 'ripple' profile due to an annulus at $50 \pm 5r_{g}$ (the dark 
blue solid curve in Fig.~\ref{fig:feka}) with the line profile due to a steadily draining 
inner disk which drains 'inside out'. The black solid curve corresponds to the dark blue solid curve in 
Fig.~\ref{fig:feka}. To mimic an 'inside out' draining inner disk, we held both the inner and outer gap edges (at 
$45,55r_{g}$ respectively) fixed, while allowing the inner disk edge ($r_{\rm inner}=6r_{g}$ for the black curve) to migrate outward (on the disk 
viscous timescale) to $10r_{g}$ (red), $20r_{g}$ (green), $30r_{g}$ (dark blue), $40r_{g}$
 (light blue). For a fiducial disk with $\alpha=0.01,(H/r)=0.1$ 
around a $10^{6}(10^{8}M_{\odot})$ supermassive black hole, we expect the progression of profiles from solid black curve to light blue curve depicted here to last approximately $\sim 0.5 (50)$ years. All curves are binned at approximately the energy resolution ($\sim 7$eV) expected for \emph{Astro-H}.
\label{fig:inside_out}}
\end{figure}

In Fig.~\ref{fig:inside_out}, we follow the draining of the inner disk due to a black hole 
migrator that has stalled at $50r_{g}$. The satellite black hole has opened a gap in the 
disk at $50\pm 5r_{g}$ We assume that the outer gap edge remains fixed at $55r_{g}$ and that the inner disk drains on the viscous disk timescale. In the time it takes the inner disk to drain, we assume that no gas enters the gap and for now we neglect the effect of 'pile-up' of gas mass at the outer edge of the gap ($55r_{g}$). The viscous disk timescale is given by \citep{arm10}
\begin{equation}
\tau_{\alpha}=\frac{1}{\alpha}\left(\frac{H}{r}\right)^{-2}\frac{1}{\omega}
\label{eq:talpha}
\end{equation}
where $\omega$ is the Keplerian angular frequency. For a puffed-up inner disk with 
$H/r \sim 0.1$ and $\alpha \sim 0.01$, the inner disk will drain in 
$\sim 10^{4} \tau_{\rm d}$ where $\tau_{\rm d}$ is the dynamical timescale at the gap 
inner edge (in this case $45r_{g}$). Note that the viscous timescale is shortest at small radii, so we expect the inner disk to drain 'inside out', i.e. $r_{\rm inner} \rightarrow r_{1}$ over time. Therefore we expect the surface density to first drop dramatically at small radii as material gets accreted. The surface density of gas in annuli at large radii in the inner disk will drop as it is 'smeared out' across interior radii as it accretes, but we shall ignore the mean emission from interior to $r_{\rm inner}$ and simply approximate the inside out drainage as a migration outward of $r_{\rm inner}$ to $r_{1}$.

Fig.~\ref{fig:inside_out} shows the effect as the inner disk drains in this simple 'inside out' manner. The black curve shows the inner disk as for the dark blue curve in Fig.~\ref{fig:feka}, i.e. annulus spanning 45-55$r_{g}$, inner disk radius at $6r_{g}$. The progression in Fig.~\ref{fig:inside_out} shows the effect of allowing the inner disk edge ($r_{\rm inner}$) to migrate \emph{outward} on the disk viscous timescale. For a stalled $q=3 \times 10^{-3}$ secondary black hole at $50r_{g}$ around a $10^{6}M_{\odot}$ ($10^{8}M_{\odot}$) supermassive black hole, the inner disk will drain on the approximate timescale $\sim 0.5(50)$yrs, far shorter than the timescale ($\tau_{\rm GW}$) for merger by gravitational wave emission. The progression from black curve to red curve ($r_{\rm inner}=10r_{g}$), to green curve ($r_{\rm inner}=20r_{g}$) occurs very quickly. The disk drainage progression in Fig.~\ref{fig:inside_out} spends most of its time between the dark blue and light blue curves. Observationally, the key characteristics to indicate this 'inside out' disk drainage are that the positions of the notches in the broad line (due to the gap) remain fixed, but the red and blue wings are suppressed monotonically as the line intensity diminishes. In this case, multiple repeated observations using the large effective area of \xmm EPIC could be used to collect flux from the red and blue wings of the broad line. Even with the limited EPIC energy resolution ($\geq 0.1$keV), the equivalent width of the line wings (as well as the notch positions) can be measured down to fractions of individual exposures ($\sim 10$ks intervals). However, we expect the monotonic decrease in the equivalent width of the wings to be most significant over longer intervals ($\sim$ yr) while the positions of the notches remains fixed. It is possible that disk drainage could occur on timescales faster than fiducial estimates of $\tau_{\alpha}$. For example, the sharp gradients at the cavity edge can cause viscous diffusion to be significantly faster than the local viscous timescale for uniform (non-piled-up) gas \citep{mp05,hkm09}. Furthermore, the effective viscosity $\alpha$ has been found to increase locally near the cavity edge by more than an order of magnitude \citep{ShiKrolik:2011,Noble+2012}; a similar increase is seen in the inner regions of single-BH disks \citep{Penna+2012}. So, even for secondaries around very massive primaries, changes in the Fe~K$\alpha$ line due to disk drainage could be detectable over a period of several years. The reverse process of gap-filling \citep{tanmen10,tan10}, starting after decoupling from the disk and lasting beyond the MBH merger, can produce a time-reversed sequence lasting$\sim 10$yrs for a $10^{6}M_{\odot}$ SMBH. 

\begin{figure}
\includegraphics[width=3.35in,height=3.35in,angle=-90]{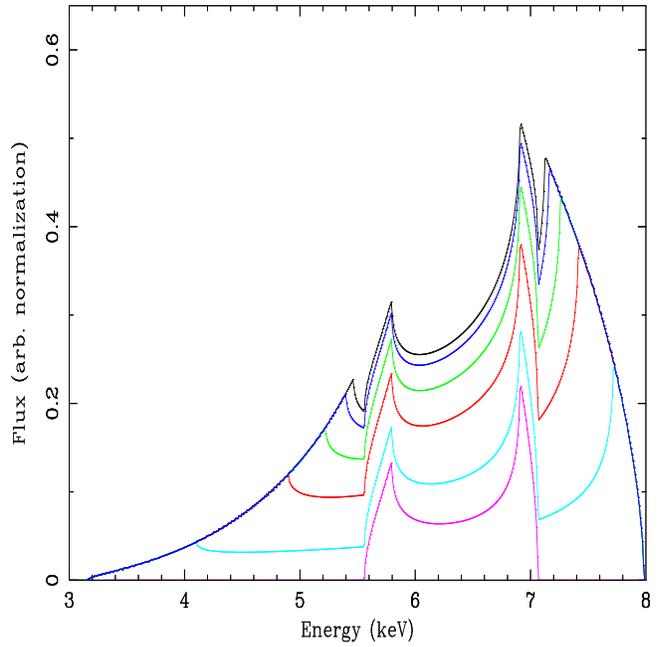}
\caption{Comparing the 'ripple' profile due to an annulus at $50 \pm 5r_{g}$ (the dark 
blue solid curve in Fig.~\ref{fig:feka}) with the line profile due to a steadily draining 
inner disk. The black solid curve corresponds to the dark blue solid curve in 
Fig.~\ref{fig:feka}. To replicate an 'outside-in' draining inner disk, we held the outer gap edge (at 
$55r_{g}$) fixed, while allowing the inner gap edge ($r_{1}$) to migrate inward (on the disk viscous timescale) to $40r_{g}$ (dark blue), $30r_{g}$ (green), $20r_{g}$ (red), $10r_{g}$ (light blue) and $0r_{g}$ (purple). This form of drainage is unlikely as the viscous disk timescale is shortest at small radii, although $\alpha$ and $H/r$ may change significantly as radii decrease. For a fiducial disk with $\alpha=0.01,(H/r)=0.1$ 
around a $10^{6}(10^{8}M_{\odot})$ supermassive black hole, we expect the progression of profiles from solid black curve to purple curve depicted here to last approximately 
$\sim 0.5 (50)$ years. All curves are binned at approximately the energy resolution ($\sim 7$eV) expected for \emph{Astro-H}.
\label{fig:drain}}
\end{figure}

If instead, we assume the drainage is 'outside in', the disk depletes first at larger radii and then at smaller radii in to $r_{ISCO}$, to form the cavity, the change in the line profile will be different. Outside-in drainage is unlikely, since the viscous timescale is shortest for small radii, although $\alpha(r)$ and $H/r$ in eqn.~\ref{eq:talpha} may change significantly with decreasing radius. Fig.~\ref{fig:drain} shows the effect of outside-in drainage, where $r_{1}$ decreases to $r_{\rm inner}$. For most of this time, the line profile will be somewhere between the black and green curves. The final stages of disk drainage will happen very fast and the rate of change in the broad FeK$\alpha$ line profile will be large (going from red curve to purple curve). As the inner gap edge decreases to $40r_{g}$ (dark blue curve), $30r_{g}$ (green curve), $20r_{g}$ (red curve), $10r_{g}$ (light blue curve) and $r_{g}$ (i.e. empty cavity; purple curve). 

\begin{figure}
\includegraphics[width=3.35in,height=3.35in,angle=-90]{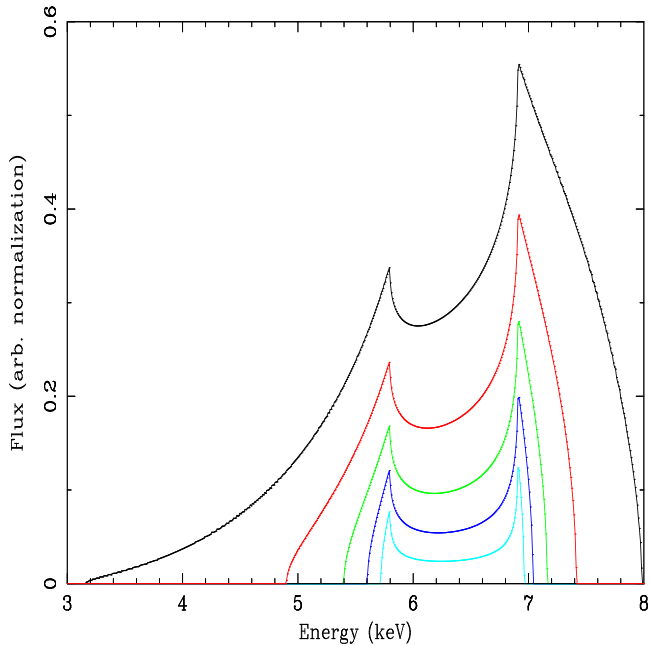}
\caption{The change in the broad FeK$\alpha$ line profile due to the draining of the 
entire inner disk due to a stalled gap-opening satellite black hole, where $r_{1}=100r_{g}$. To replicate a draining inner disk, we assumed the broad FeK$\alpha$ emission 
originates entirely from within $100r_{g}$, with $\theta=60^{o}$ and $r^{-2.5}$ X-ray 
emissivity profile as assumed above. We do not see 'notches' as in Fig.~\ref{fig:inside_out} since the gap does not imprint itself on the inner disk $<100r_{g}$. The solid black curve corresponds to the broad 
FeK$\alpha$ profile from the inner disk extending from $100r_{g}$ inward to $6r_{g}$. 
Since we expect the disk viscous timescale to increase with radius, we assume the disk drains in an 'inside out' manner, so $r_{\rm inner} \rightarrow r_{1}$ and we ignore emission from gas 'smeared out' within $r_{\rm inner}$. As the inner edge of the disk increases radially to $20r_{g}$ (red curve), $40r_{g}$ (green 
curve), $60r_{g}$ (dark blue curve), $80r_{g}$ (light blue curve), the broad component of 
the line decreases in magnitude and the blue and red wings are increasingly suppressed. For a fiducial disk with $\alpha=0.01,(H/r)=0.1$ around a $10^{6}(10^{8}M_{\odot})$ supermassive black hole, we expect the progression of profiles from solid black curve to light blue curve depicted here to last approximately $\sim 1.5(150)$yrs. Two-thirds of this time is taken up with the progression from green curve to light-blue curve. All curves are binned at approximately the energy resolution ($\sim 7$eV) expected for \emph{Astro-H}.
\label{fig:drain_ann100}}
\end{figure}

If the gap-opening black hole stalls further out, say at $\geq 100r_{g}$ or 
greater as might be expected \citep{hkm09}, the inner disk will drain on the viscous timescale of the inner edge of the gap at $\geq 100r_{g}$, while the outer disk is dammed. In this case, for a $q=3 \times 10^{-3}$ gap-opening black
 hole stalled such that $r_{1}=100r_{g}$ in a $\alpha=0.01,H/r=0.1$ gas disk around a 
$10^{6}M_{\odot}(10^{8}M_{\odot})$ supermassive black hole, 
the inner disk will drain in $\sim 1.5(150)$yrs $\ll \tau_{\rm GW}$. If the disk drains 'inside out' ($r_{\rm inner} \rightarrow r_{1}$ over time), as we expect since the viscous timescale is shortest at small radii, the broad line component over that time will follow the progression in Fig.~\ref{fig:drain_ann100}, assuming that disk gas at $\geq 100r_{g}$ is held up by the 'dam wall'. In our example of a $q=3\times 10^{-3}$ secondary around a $10^{6}(10^{8})M_{\odot}$ primary, the line profile will change from the black to green curve over $0.5(50)$years and from green curve to light-blue curve in $1(100)$years. If the inner disk is draining 'inside out' in this way, the key observational characteristic to search for is the increasing suppression of the red and blue wings of the line over time. This model can be tested using archival and ongoing observations with the large effective area of \xmm EPIC. In this case, the binned-up wings of the broad component will decrease in a consistent monotonic pattern over time (quickly at first, then more slowly). If instead, the $\alpha$ and $H/r$ profiles of the disk mean that inside-out drainage does not occur, and an 'outside-in' disk drainage somehow manages to happen, the red and blue 'horns' of the line profile will ripple outward from the line centroid energy as the overall broad component intensity diminishes. In this case, the low energy resolution of \xmm can capture the migration of the horns red-ward and blue-ward, but only over long timescales. The higher energy resolution of \emph{Astro-H} and \emph{IXO/Athena} could resolve horn migration over much shorter timescales. 

Note that if the X-ray continuum originates in a hot corona concentrated mostly above the innermost disk then during and after cavity formation, we should expect some fraction of the overall X-ray continuum to decline. If the inner disk drains 'inside out' to form a cavity, we should expect a rapid drop of the X-ray continuum together with a strong decrease in broad line intensity (most pronounced in the wings). Observing such an effect would allow us not only to test models of massive mergers, but also allow us to constrain the fraction of the X-ray continuum that originates in the central $\sim 100r_{g}$. If the disk drains 'outside in' (much less likely), we expect the drop in the continuum to match the rippling apart of the horns of the Fe~K$\alpha$ line.

\subsection{Detecting pile-up in AGN disks}
\label{sec:dam}
In the discussion of disk drainage above, we ignored the effect of the continuing inflow of gas from the outer disk. As the migrating secondary stalls and the disk interior to it drains on the viscous timescale, mass is still flowing inward in the disk. Gas should build up at the outer gap or cavity edge on the viscous timescale ($\tau_{\alpha}$) at that disk radius \citep{sc95,ivan99,Kocsis+2012a}. We can think of the outer gap radius or cavity edge as a dam holding back the inflow of gas. Here we discuss how damming inflowing gas at the outer wall of the gap or cavity has an effect on the observed broad Fe~K$\alpha$ line component.

We created a naive toy model to represent the damming of inflowing gas at the outer gap wall and we considered two simple cases. First, as the inner disk drains onto the primary, creating a cavity in the inner disk (as in Figs.~\ref{fig:drain} and ~\ref{fig:drain_ann100} in \S\ref{sec:cavities}), gas piles up at the outer gap wall. Second, gas piles up at the edge of a pre-existing cavity. In both cases our toy model assumes a simple, uniform  enhancement of disk emission in a annulus at the disk truncation radius. Pile-up behind the dam is unlikely to be uniform in reality, particularly for large values of $q$. A dense tidal bulge tracking the secondary can form on the dam wall \citep{ShiKrolik:2011,roedig11,Noble+2012,dzm12}, which will orbit at the orbital time for the cavity wall. This can lead to an oscillation in the effect discussed below but we shall return to this in future work. In order to move the stalled secondary, we should expect a pile-up comparable to the mass of the secondary. We translated this into a simple density enhancement over a standard thin disk of a factor $\sim2-5$ distributed uniformly within $\sim 10\%$ of the cavity edge. This is roughly consistent with profiles of the pile-up seen in simulations \citep[e.g.][]{mm08,Cuadra:2009}, although over time we might expect a $\Sigma(r) \propto r^{-1}$ build up to large radii behind the dam \citep{Kocsis+2012a}. We shall investigate the effects of different types of 'damming' in future work. Here we are not attempting to reproduce details of the damming; we neglect disk heating, pressure and expansion. Rather, we simply want to understand the sense in which the broad Fe~K$\alpha$ line component changes when we enhance the Fe~K$\alpha$ emission at the disk edge in a naive manner.  

\begin{figure}
\includegraphics[width=3.35in,height=3.35in,angle=-90]{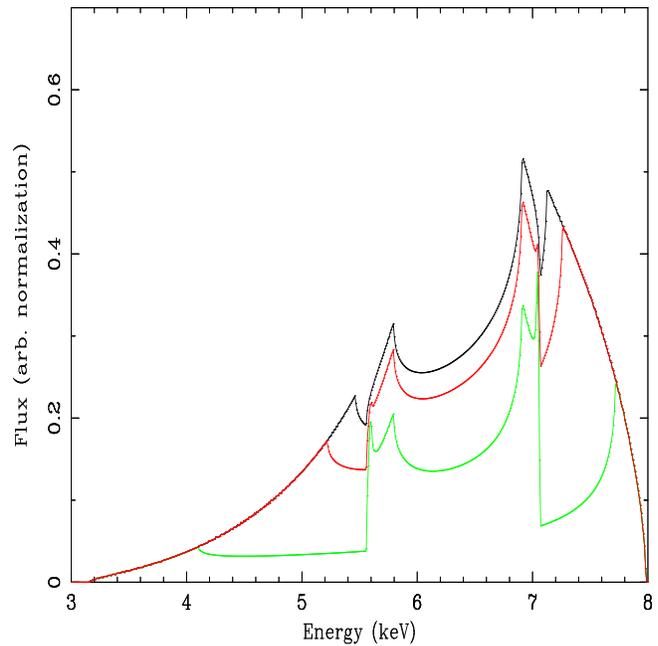}
\caption{The change in the broad FeK$\alpha$ line profile due to draining of the 
entire inner disk ('outside-in') due to a stalled migrating satellite black hole at $50r_{g}$, while gas piles up at the gap outer edge. The solid black curve corresponds to the broad 
FeK$\alpha$ profile due to the inner disk with a cavity spanning $45-55r_{g}$ (same as black curve in Fig.~\ref{fig:drain}). The red curve corresponds to $r_{1}=30r_{g}, r_{2}=55r_{g}$ with an enhancement of emission by $\times 2$ in the annulus spanning $55-60r_{g}$. The green curve corresponds to $r_{1}=10r_{g}, r_{2}=55r_{g}$ with an enhancement of emission by $\times 4$ in the annulus spanning $55-60r_{g}$. All curves are binned at approximately the energy resolution ($\sim 7$eV) expected for \emph{Astro-H}.
\label{fig:drain_disk_cav_dam}}
\end{figure}

In the first case, where the inner disk at $<r_{1}$ is draining onto the primary as in Fig.~\ref{fig:drain}, we assume that there is an enhancement of the emission from gas  in an annulus immediately outside $r_{2}$. Fig.~\ref{fig:drain_disk_cav_dam} shows the inner disk draining 'outside in', but now we uniformly  enhance  Fe~K$\alpha$ emission in an annulus spanning $55-60r_{g}$ immediately outside $r_{2}$ to model pile-up. As the inner gap edge ($r_{1}$) decreases to $30r_{g}$ (red curve), we add a uniform (arbitrary) $\times 2$ enhancement of the Fe~K$\alpha$ emission from disk radii spanning $55-60r_{g}$. As the inner gap edge decreases to $r_{1}=10r_{g}$ (green curve), the Fe~K$\alpha$ emission from disk radii spanning $55-60r_{g}$ is enhanced by a uniform (arbitrary) $\times 4$ factor. By comparing Fig.~\ref{fig:drain_disk_cav_dam} with Fig.~\ref{fig:drain}, we can see that substantial damming of the outer disk has a noticeable effect on the broad component of the Fe~K$\alpha$ line. In particular we see that the 'horns' of the broad component each become double-pronged as the 'pile-up' beyond $r_{2}$ increases. Observationally, we should expect the double-pronged horns to persist until the dam bursts. If the dam burst is sudden and catastrophic, we expect the extra horns in the line profile will ripple outward to redder and bluer energies respectively as the 'overdense' material collapses inward. If instead, there is a slow leak from the dam, the flux in the extra horns in the line profile will decrease and become dispersed to energies redward and blueward of the horns.

\begin{figure}
\includegraphics[width=3.35in,height=3.35in,angle=-90]{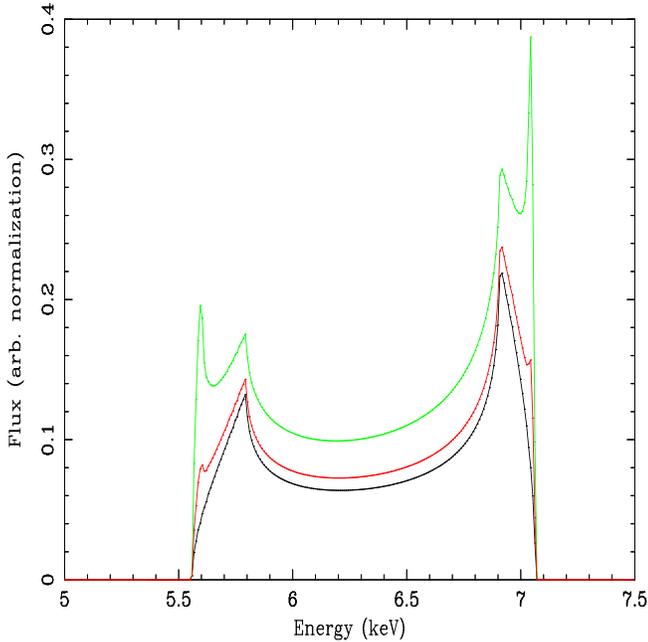}
\caption{The change in the broad FeK$\alpha$ line profile due an inner disk truncated by a cavity at $r_{1}=r_{2}=55r_{g}$, while gas piles up at the gap outer edge. The solid black curve corresponds to the broad FeK$\alpha$ profile due to the inner disk truncated by a cavity at $<55r_{g}$ (same as purple curve in Fig.~\ref{fig:drain}). The red curve corresponds to an enhancement of emission by $\times 2$ in the annulus spanning $55-60r_{g}$. The green curve corresponds to enhancement of emission by $\times 5$ in the annulus spanning $55-60r_{g}$. All curves are binned at approximately the energy resolution ($\sim 7$eV) expected for \emph{Astro-H}.
\label{fig:dam_disk_cav}}
\end{figure}

In the second case, we assume that a cavity has been excavated, truncating the disk at $r_{1}=r_{2}=55r_{g}$. We also assume the outer disk is damming inflowing gas in an annulus spanning $55-60r_{g}$. Fig.~\ref{fig:dam_disk_cav} shows the effect over time of uniform pile-up at the edge of the cavity ($r_{2}$). The black curve shows the broad line from the region outside the cavity ($>55r_{g}$), with no damming of the accretion flow (equivalent to the purple curve in Fig.~\ref{fig:drain}). The red curve shows an enhancement of $\times 2$ over 'normal' emission in an annulus spanning $55-60r_{g}$. The green curve shows an enhancement of $\times 5$ over 'normal' emission in the annulus at $55-60r_{g}$. From Fig.~\ref{fig:dam_disk_cav}, the key observational characteristic of a 'damming' of the accretion flow is a double-horned broad Fe~K$\alpha$ line component (as in Fig.~\ref{fig:drain_disk_cav_dam}). This is because the broadest component in the observable Fe~K line complex is emitting substantially more than the surrounding disk. In this case, the blue- and red-shifted horns of the line from the innermost annulus stand out relative to the blue- and red-shifted horns of the line from annuli further out in the disk. At some point, when the pressure is large enough, the dam will overflow. If the overflow is continuous, we expect a dynamic equilibrium configuration of 'double-peaked' horns, where flux lost from the annulus is replaced by inflowing material. If the overflow is sudden, the outermost of the double-peaked horns will decay rapidly as flux gets redistributed blue-ward and red-ward of the horns. Thus, if we observe a broad Fe~K component with double-peaked horns that \emph{do not} change over the viscous timescale, there is a dynamic equilibrium between pile-up behind and leakage from the dam. The energy resolution of \emph{Astro-H} is necessary to begin to detect the 'double-peaked' horns corresponding to 'dammed' accretion disks. However, to systematically study such effects (rather than rely on serendipity), future proposed missions will require much larger effective area than even that planned for \emph{IXO/Athena}.

\subsection{Detecting a disk around the secondary black hole}
\label{sec:minidisk}
In our discussion of gaps and cavities above, we have ignored accretion onto the secondary. However (see \S~\ref{sec:gaps}), it is possible that the secondary's accretion disk persists long after the inner disk has drained. An accretion disk within the Hill radius $R_{\rm H} =(q/3)^{1/3}R$ of a $q=0.03$ secondary located at $50r_{g}$ in a cavity in the innermost AGN disk, has a radius of $11r_{g}$ in units of the primary black hole mass or $365r_{g,s}$, where $r_{g,s}$ corresponds to gravitational radii in units of the secondary black hole mass. The viscous timescale in the disk around the secondary is $\tau_{\alpha,s}=(1/\alpha_{s})(H_{s}/r_{s})^{-2} 1/\omega_{s}$ where the parameters are as in eqn.(~\ref{eq:talpha}), but the s-subscript refers to the accretion disk around the secondary (which may have different properties to the main AGN disk around the primary). Assuming $\alpha_{s}\sim 0.01, H_{s}/r_{s} \sim 0.1$, $\tau_{\alpha,s} \sim 10^{4} \tau_{\rm d,s}$, where $\tau_{\rm d,s}$ is the dynamical timescale in the secondary disk, a disk extending to $365r_{g,s}$ around a $q=0.03$ secondary orbiting a $10^{6(8)}M_{\odot}$ supermassive primary black hole, will drain in $\sim 2(200)$yrs. Periodic dam overflows or continuous dam leakage could keep the disk persisting (see \S\ref{sec:gaps}) around the secondary long after a cavity has formed. There could also be a small remaining inner disk around the primary in this case, which will add flux to the red and blue wings of the Fe complex. Since motion will be about the center of mass of the merging binary, a small disk component around the primary will wobble, which may be detectable for large mass secondaries ($q \geq 0.01$) with the energy resolution of future missions such as \emph{Astro-H} and \emph{IXO/Athena}, however, we shall leave discussion of the effect of oscillations of the primary component to future work. In the following  discussion, we consider only flux from simple accretion disks around the secondary and ignore the flux contribution from the gas streamers within the cavity. 

The secondary disk will add an additional component to the observed broad Fe~K$\alpha$ line as it orbits the primary. The observational effect will be most obvious at energies red-ward and blue-ward of the line component produced by the main AGN disk outside the cavity (e.g. the purple curve in Fig.~\ref{fig:drain}). A key difference between the ripple effect due to a disk gap, and the effect due to a mini-disk around the secondary, is that the former occurs on the AGN disk viscous timescale and the latter occurs on the much shorter orbital timescale. Thus, the effect consists of adding a secondary, low intensity broad Fe~K$\alpha$ component that oscillates across the primary broad Fe~K$\alpha$ component on the orbital timescale. \emph{The secondary line component oscillates approximately between the red and blue horns of the primary line component over the orbital timescale}. For a secondary located at $50r_{g}$ from a $10^{6(8)}M_{\odot}$ primary, one complete cycle of oscillation of the line centroid of the secondary component between the red and blue horns of the primary component will take $\sim 2\pi R_{s}/v_{s} \sim 10^{4(6)}$s where $R_{s}=50r_{g}$ is the secondary orbital radius and $v_{s}$ is the secondary orbital velocity. 

\begin{figure}
\includegraphics[width=3.35in,height=3.35in,angle=-90]{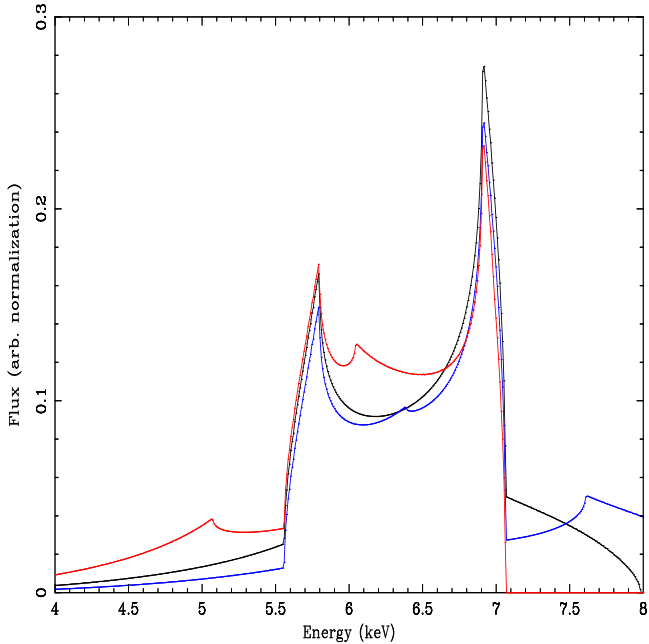}
\caption{The change in the observed broad FeK$\alpha$ line profile due to an accretion disk around the secondary black hole located at $50r_{g}$. The black curve corresponds to the Fe~K$\alpha$ emission from the main AGN disk ($55-100r_{g}$) plus a weak secondary broad component ($10\%$ of the intensity of the full disk profile) due to an accretion disk around the secondary black hole centered on the line centroid energy (6.40keV). The red curve shows the effect of shifting the centroid of the weak secondary component red-ward to $5.60$keV. The blue curve shows the effect of shifting the centroid of the weak secondary component blue-ward to $7.05$keV. The progression from red curve to blue curve occurs over half the orbital time of the secondary. Observationally, we expect a 'see-saw' oscillation between the blue and red wings of the line as the secondary accretion disk orbits the primary black hole inside the circumbinary disk cavity. All curves are binned at approximately the energy resolution ($\sim 7$eV) expected for \emph{Astro-H}.
\label{fig:2nddisk}}
\end{figure}

In Fig.~\ref{fig:2nddisk} we show the effect of adding an accretion disk around a secondary black hole located at $50r_{g}$ in a cavity which truncates the AGN disk at $55r_{g}$. We assume that the circumsecondary disk properties are identical to the circumbinary disk properties (emissivity, disk inclination) except for the cavity. We also assume that the secondary broad line flux is $10\%$ of the primary flux \emph{without a cavity} (the black curve in Fig.~\ref{fig:drain_ann100}).  In Fig.~\ref{fig:2nddisk}, the black solid curve corresponds to the superposition of the broad line profile from a AGN disk with cavity (purple curve in Fig.~\ref{fig:drain}), with a secondary broad line component centered on the line centroid energy (6.40keV). The red-curve in Fig.~\ref{fig:2nddisk} shows the AGN disk/cavity profile with the secondary disk line profile superimposed at the most red-shifted part of its orbit (line centroid is $\sim 5.6$keV). The blue curve in Fig.~\ref{fig:2nddisk} shows the AGN disk/cavity line profile, with the secondary disk line profile superimposed at the most blue-shifted part of its orbit (line centroid is $\sim 7.05$keV). Thus, Fig.~\ref{fig:2nddisk} shows the effect of one half of the orbit, as the secondary component centroid moves between the red horn to the blue horn of the primary component. The key observable for detecting disks around secondary, satellite black holes is therefore a periodic see-saw oscillation between the red and blue curves as in Fig.~\ref{fig:2nddisk}. If there were a weak circumprimary disk (say $\leq 10r_{g}$), it would add a very broad component to the red and blue wings of the Fe~K line complex. Observationally this would be equivalent to adding a component to the red and blue wings that will vary on the timescale of the primary wobble around the binary center-of-mass. The wobble of the primary component depends on the mass ratio ($q$) of primary to secondary and may be detectable with the energy resolution of \emph{IXO/Athena} in very long exposures.  Here we shall focus on oscillations due to the secondary disk and we shall leave discussion of the effect of oscillations of the primary component to future work.

\begin{figure}
\includegraphics[width=3.35in,height=3.35in,angle=-90]{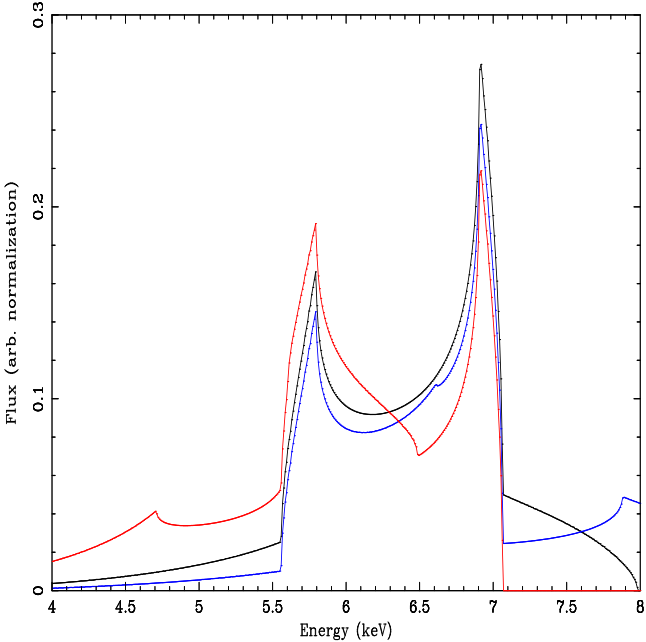}
\caption{The change in the observed broad FeK$\alpha$ line profile due to an accretion disk around the secondary black hole as in Fig.~\ref{fig:2nddisk}, except the secondary is located closer in, at $30r_{g}$. The black curve corresponds to the Fe~K$\alpha$ emission from the main AGN disk ($55-100r_{g}$) plus a weak secondary broad component ($10\%$ of the intensity of the full disk profile) due to an accretion disk around the secondary black hole centered on the line centroid energy (6.40keV). The red curve shows the effect of shifting the centroid of the weak secondary component red-ward to $5.2$keV. The blue curve shows the effect of shifting the centroid of the weak secondary component blue-ward to $7.3$keV. The progression from red curve to blue curve occurs over half the orbital time of the secondary. Observationally, we expect a 'see-saw' oscillation between the blue and red wings of the line as the secondary accretion disk orbits the primary. All curves are binned at approximately the energy resolution ($\sim 7$eV) expected for \emph{Astro-H}.
\label{fig:2nddisk30rg}}
\end{figure}

In Fig.~\ref{fig:2nddisk30rg} we show the same effect, this time for a secondary located at $30r_{g}$ inside a cavity that truncates the disk at $55r_{g}$. In this case, we see a much more pronounced oscillation between the red and blue horns of the total observed broad line component. Here, the line centroid energy of the secondary oscillates between $5.2$keV and $7.3$keV. This oscillation can occur on the very short secondary orbital timescales (few-10s of ks in AGN with a primary $10^{6-8}M_{\odot}$) and could easily be detected during an extended observation with \emph{Astro-H}. Thus, the deeper the secondary accretion disk in the cavity, the easier it is to observe the oscillation effect. Fig.~\ref{fig:2nddisk30rg} looks quite different from the results of \citet{srrd12} (their Fig.~10), where there are two \textsc{laor} model disk components, corresponding to a $10^{9}M_{\odot}$ total-mass binary, with the secondary blue-shifted by $10^{4} \rm{km} \rm{s}^{-1}$ (corresponding to a separation of $\sim 10^{3}r_{g}$), much larger than the separations discussed here. The apparent  difference is mostly driven by the lack of horns in the profiles in \citet{srrd12} which is due to their choice of steeper emissivity profile $r^{-3}$ (see e.g. Fig.10 in \citep{reynow03} for the impact of this choice on broad line profile).

\begin{figure}
\includegraphics[width=3.35in,height=3.35in,angle=-90]{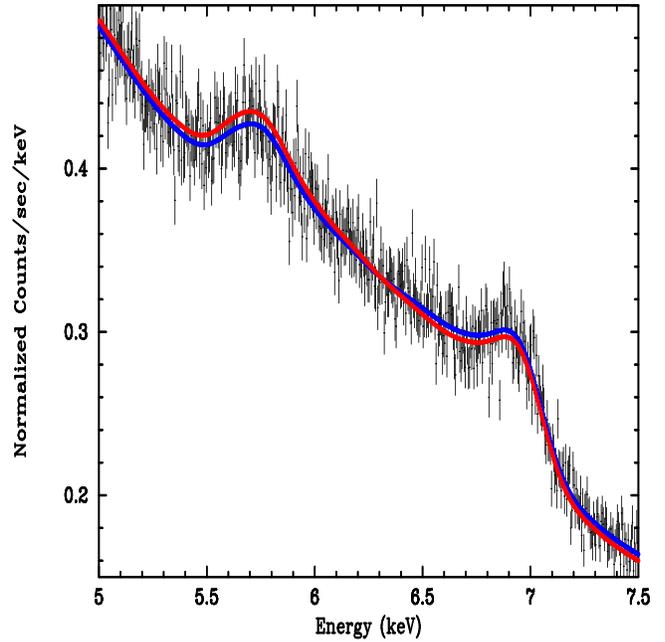}
\caption{A simulated 350ks observation with \xmm \textsc{EPIC pn} of an AGN at $z=0.01$, with a 2-10keV count-rate of $3.5 \times 10^{-11} \rm{phot} \rm {cm}^{-2} \rm{s}^{-1} $, which is best fit with a two component diskline model (assuming the narrow FeK$\alpha$ component has been subtracted). We used a simple XSPEC model \textsc{zphabs(zpowerlw+diskline+diskline)} where the absorbing Hydrogen column was $10^{21}\rm{cm}^{-2}$ and the powerlaw index was $-1.8$. The primary diskline has line centroid at 6.40keV, emissivity $k=-2.5$, $\theta=60^{o}$ and originates from $50-100r_{g}$ in a truncated inner disk, with line normalization set to $10^{-2}$ of the continuum. We ignore the narrow Fe~K$\alpha$ line component for clarity.The secondary diskline has similar parameters to the primary, but relative normalization of $0.2$ in order to highlight the sense in which the oscillation can change the best-fit. Black data points correspond to simulated data where the secondary line centroid lies at 7.3keV and the blue solid line corresponds to the best-fit model to the data. The red solid line corresponds to the best model fit when the line centroid of the secondary lies instead at 5.2keV (simulated data not shown for clarity). 
\label{fig:sim_xmm}}
\end{figure}

In Fig.~\ref{fig:sim_xmm} we simulate\footnote{http://heasarc.gsfc.nasa.gov/cgi-bin/webspec} a $350$ks observation with \xmm \textsc{EPIC-pn} of an AGN at $z=0.01$ with a 2-10keV photon flux of $3.5 \times 10^{-11} \rm{phot}  \rm{cm^{-2}} \rm{s^{-1}}$. We ignore the narrow line component in the model in order to illustrate the difference in best-fits to observations; in practice this model component would need to be fit to the data too. The \textsc{XSPEC} toy model was simply \textsc{zphabs(zpowerlw+diskline+diskline)} where the absorbing Hydrogen column was $10^{21}\rm{cm}^{-2}$ and the powerlaw index was $-1.8$. The primary diskline was centered at 6.40keV and due to a truncated inner disk from 50-100$r_{g}$, with $\theta=60^{o}$ and emissivity $k=-2.5$ as above, and a line normalization of $10^{-2}$ with respect to the continuum. The secondary diskline had similar parameters to the primary diskline except it ran from 6-100$r_{g,s}$, where $r_{g,s}$ represents  gravitational radii of the secondary. We set the normalization of the secondary diskline to $\sim 0.2$ of the primary to illustrate the difference between the best fits to two simulated observations. The simulated data in in Fig.~\ref{fig:sim_xmm} corresponds to line centroid of the secondary at 7.3keV and the blue solid line is the best model fit to the data. The red solid line in Fig.~\ref{fig:sim_xmm} corresponds to the best model fit when the line centroid of the secondary lies at 5.2keV, although we do not show the simulated data in the latter case for ease of presentation. Flux differences in the red wing between the two models in Fig.~\ref{fig:sim_xmm} differ by a few percent and the key observational test in this case is that the best-fit model must oscillate between the blue and red curves on a regular orbital timescale. As we can see, even in two long simulated exposures with \xmm and a relatively intense secondary disk component, it is difficult to disentangle the secondary disk component, even if we bin up the data. However, repeated observations of bright AGN can build up statistical significance in the red and blue wings, allowing us to track the oscillation of the secondary over time. Thus, observations of AGN in the Fe~K band with \xmm \textsc{EPIC-pn} should be searched for serendipitous oscillations. A systematic study of models of secondary disks in AGN could be carried out with \emph{LOFT} since the proposed effective area ($10m^{2}$) is an order of magnitude larger than \xmm and the spectral resolution ($\sim 0.25$keV) is sufficient for coarse binning of the red and blue wings of the broad Fe K$\alpha$ line. For disentangling competing effects, ideally a Large Area High Resolution (\emph{LAHR}) telescope should be used, with effective area greater than that of \emph{LOFT} \emph{together} with the proposed spectral resolution of \emph{IXO/Athena}.

\section{Multimessenger astronomy with gravitational waves and the ripple effect }
\label{sec:gw}
Observations of a ripple effect or periodic oscillations in the broad FeK$\alpha$ line will be an excellent leading
indicator of gravitational radiation which can be detected directly independently. 
Depending on the orbital period of the binary, the GWs emitted by a supermassive black hole binary 
are detectable with pulsar timing\footnote{\url{http://www.ipta4gw.org/}} or a space-based laser-interferometric instrument like 
LISA\footnote{\url{http://lisa.nasa.gov/}} or NGO\footnote{\url{http://elisa-ngo.org/}}.
The GW measurement can constrain the source parameters completely independently 
of the electromagnetic signal with different systematics. Furthermore, the GW measurement may independently indicate the presence of ambient gas around the binary as discussed below. A serendipitous binary detection 
in one channel can be used to do a targeted search in the other for verification \citep{khm08}. 
The comparison of these measurements can provide a consistency check 
for the central mass, inclination, and orbital radius. The combination of the electromagnetic and GW measurements may allow us to constrain the geometry of the accretion disk and the binary. 
In this section we discuss the gravitational wave signatures 
that may be coincident with the electromagnetic signatures. 

The GW strain amplitude generated by a circular\footnote{ The GW waveform for general eccentric orbits 
is similar, but additional terms involving $\dot\phi$ appear in that case \citep{pm63}. } 
binary of separation $r$,  distance $D$, 
inclination $\iota$ relative to the line of sight is
\begin{equation}\label{e:GW}
\left(
\begin{array}{l}
h_+(t)\\
h_\times(t)
\end{array}
\right)
=
2 \frac{m_1 m_2}{r D}\left(
\begin{array}{l}
(1+\cos^2 \iota) \sin 2\phi(t)\\
2 \cos \iota \cos 2\phi(t)
\end{array}
\right)
\end{equation}
where $\phi(t)$ is the orbital phase along the orbit.
Here, the separation $r$ changes in a characteristic way due to GW losses and gas effects, 
which causes changes in the orbital frequency, $d \phi/dt$. The corresponding phase shift can be used to 
measure the binary parameters including the separation, eccentricity, masses, and spins \citep{cf94}
and the torques exerted on the binary by the accretion disk \citep{kyl11,yklh11,hytm12}. 
The distance to the source and the binary inclination then follows from the amplitudes of the two 
GW polarizations $h_+$ and $h_{\times}$. The sky position of the source can be inferred by measuring the GW
strain coincidentally over multiple baselines.

\subsection{PTA detections}
Pulsar timing arrays (PTAs) can measure GWs of periods weeks to years (or frequencies 
between nHz to mHz) by detecting correlated variations in the time-of-arrivals (TOAs) 
of pulses in millisecond pulsars in the Galaxy.  At these periods the binaries are relatively 
far from merger, so only four quantities are measurable for most of the sources: 
the GW amplitudes of the two polarizations (see Eq.~\ref{e:GW}), the orbital period, 
and the eccentricity, but not the inspiral rate or $\dot \phi$. However, the inspiral rate
may be measurable for some sources \citep{lee11}. In the case of PTA detections, we expect to detect the binary when it is close to decoupling from the disk \citep{sksd11} and therefore the signature of a gap or cavity in the broad Fe~K$\alpha$ line caused by disk drainage (outlined above) should be clearest. We expect to detect a few such binaries with PTAs out to $z<1$.

These observations are mostly sensitive to the total stochastic background generated by many 
supermassive black hole binaries across the Universe at low frequencies \citep{p01,svc08}. 
However the GWs of massive nearby binaries (e.g. $m_1\sim m_2 \gtrsim 10^9\,M_{\odot}$ 
within $z\lesssim 1$--2) are expected to rise well above the background level and may be 
individually resolved \citep{svv09,ks11}. The individually detectable binaries are expected to be orbiting at separations $100-200r_{g}$, where gas-driven migration is subdominant relative to gravitational wave losses \citep{ks11}. These GW observations would be able to localize 
such sources on the sky by comparing the TOAs of several pulsars in different directions. 
The sky localization and inclination measurement accuracy are of order degrees 
\citep{sa10,lee11,bs12,petiteau12}. 

The individually resolvable massive binaries have a typically comparable mass-ratio ($q \sim 1$). These
binaries are expected to open a large cavity in the disk. The FeK$\alpha$ profiles may show
a double structure if the non-axisymmetric streams can penetrate the gap and create an  accretion
disk around the binary components \citep{srrd12}. These binaries may also exhibit periodic 
electromagnetic variability \citep{mm08,hkm09}. With large cavities expected to be excavated in the inner disk, we expect to simultaneously detect broad Fe~K$\alpha$ profiles in these AGN similar to the individual curves shown in \S~\ref{sec:vars_models}. If we are lucky enough to detect a massive binary via PTA as the cavity is forming, we will simultaneously observe the variable broad Fe~K$\alpha$ profiles, together with ripple effects and oscillations outlined in \S~\ref{sec:vars_models}.

Since the inspiral rate is typically not available for these sources, one cannot directly measure 
the effects of gas for individual binaries from PTA observations. Gas driven migration 
may leave an imprint on the stochastic GW background by reducing the number of binaries 
in the Universe at large separations \citep{ks11}. For individual binaries, the effects of an accretion disk 
may be inferred from the GW signal indirectly by measuring a large orbital eccentricity and 
the absence of spin-orbit precession. The angular momentum exchange with an accretion 
disk leads to the excitation of orbital eccentricity for widely separated binaries if a cavity forms 
in the disk, while the eccentricity is predicted to quickly decrease in the absence of an accretion disk 
or during the later stages of the inspiral due to GW emission. However, the eccentricity is 
also increased by scattering of stars \citep{amf09,s10}, albeit to a smaller level. 
The eccentricity may be detected with PTAs \citep{en07}.
SMBH binaries with a large orbital eccentricity may be indicative of an accretion disk. 
Secondly, spin-orbit precession may also be detectable if the spins are not aligned with the orbital plane 
\citep{mingarelli12}. Such a configuration would argue against the presence of a gas disk, which tends to align the binary spins \citep{bog07}.

\subsection{LISA/NGO detections}
LISA/NGO will be sensitive to much smaller periods between 10 hours to seconds,
which corresponds to frequencies from $3\times 10^{-5}\,$ to 1\,Hz and SMBH 
masses of $M\sim 10^5$--$10^7\,M_{\odot}$ at separations from $\sim100\,r_g$ 
down to merger. In this frequency range, the inspiral of SMBH binaries is much more rapid 
and the astrophysical stochastic GW background is much more quiet, dominated by Galactic 
white dwarf binaries \citep{nyp01}. In this case, for a massive binary very close to merger, if an annulus is maintained in the disk, we expect to simultaneously observe a broad Fe~K$\alpha$ component in the late stage of the ripple effect (e.g. the red curve in Fig.~\ref{fig:feka}). If instead, the binary merger is occurring within a cavity, we expect a strongly dammed cavity inner wall, and a simultaneous Fe~K$\alpha$ broad line profile similar to the green curve in Fig.~\ref{fig:dam_disk_cav}.

 The GW signal of an IMBH or an SMBH spiraling 
into a SMBH is much stronger than the background level, and may be detected 
to high significance to high redshifts ($z\sim 10$ for comparable mass ratio binaries, 
\citealt{h02,ngo}). A sensitive measurement of the inspiral waveform for the two GW 
polarizations can be used to infer the binary masses and orbital parameters to a 
relative accuracy better than $10^{-3}$ for a signal to noise of 30 \citep{bc04}.  
The source localization accuracy with LISA/NGO is typically a few degrees several 
months before merger \citep{khm08}. This improves to a few tens of arcminutes for comparable--mass 
spinning binaries approaching the ISCO at $z\sim 1$ if the spins are not aligned with the disk 
\citep{lh08,McWilliams11}. If the disk plane and the binary spin align \citep{bog07}, there is no 
spin-orbit precession feature in the GW signal, and the binary localization 
becomes worse by a factor of a few \citep{lhc11}. This error volume however is still sufficiently small to allow a unique identification of an AGN counterpart \citep{kfhm06}.

Due to the high signal to noise ratio and the large number of orbital cycles observable 
with LISA/NGO ($N_{\rm cyc} \gtrsim 10^4$), the imprints of gas effects may be detected 
directly by measuring the corresponding perturbations of the GW phase \citep{kyl11,yklh11}. 
The largest perturbation is due to the spiral density waves excited by the binary which 
removes angular momentum from the binary analogous to planetary migration. Further 
perturbations arise due to the hydrodynamic drag on the secondary and the mass increase 
of the secondary due to accretion. These effects are more significant for smaller 
secondary masses. For an extreme mass ratio binary, $m_1/m_2 \lesssim 10^{-4}$, 
the GW phase shift due to the spiral waves are expected to be measurable with LISA 
during the last year of inspiral \citep{kyl11,yklh11}. The GW measurements may be 
directly sensitive to the presence of a gap since a gap affects the gaseous torques acting 
on the binary, and hence, the GW phase. 

For comparable mass-ratio binaries, gas effects are much less 
significant, only detectable by second generation space-based instruments
like Decigo or BBO \citep{hytm12}. In this case, the merger may take place 
in a central empty cavity in the disk where the gaseous torques are negligible. 
However, eccentricity of order $e=10^-3$ to $10^{-2}$ in LISA/NGO detections 
may be an indirect signature of an earlier coupling with a massive circumbinary 
disk  \citep{an05,roedig11}. LISA/NGO will allow to measure the eccentricity 
to a much higher accuracy $10^{-4}$ \citep{bc04,mkfv12}.

\section{Conclusions and Future Work}
\label{sec:conclusions}
If a gap-opening, migrating black hole ends up in the innermost regions of an AGN disk, analogous to a gap-opening 'hot Jupiter' in a protoplanetary disk, a unique and 
predictable pattern of variability appears in the broad component of the Fe~K$\alpha$ 
line. The ripple effect and oscillations in the Fe~K$\alpha$ line outlined in this paper are potentially detectable in long exposures with future missions such as \emph{Astro-H}, \emph{IXO/Athena} and \emph{LOFT}. Oscillations may be detectable in long, repeated exposures with \xmm \textsc{EPIC-pn}, such as may be found for popular sources in the archive. Here we have shown the range of ripple effects observable as a function of disk and merger properties, including inclination to observers' sightline, gap-width (or cavity size), disk emissivity profile, damming of the accretion flow and a mini-disk around the secondary black hole. 

Detection of a ripple effect or periodic oscillations in the broad component of Fe~K$\alpha$ from an AGN will provide advance warning of gravitational waves due to an impending merger in this AGN. For example, an observation of a broad FeK$\alpha$ line ripple profile given by the dashed red-curve in Fig.~\ref{fig:feka} from a supermassive black hole of mass $\sim 10^{6}M_{\odot}$ predicts a merger event in this source within a year. Once gravitational waves consistent 
with a binary black hole merger are detected, an archival search for a ripple effect or oscillations in a broad FeK$\alpha$ line will help localize the gravitational wave detection. Departures from the predicted ripple effect in the final stages of merger will allow us to test the predictions of strong gravity \emph{independent} of the detection of gravitational radiation.
  
\section*{Acknowledgements}
We acknowledge very useful discussions with Cl\'{e}ment Baruteau, Laura Blecha,  Massimo Dotti, Frits Paerels, Alberto Sesana and Tahir Yaqoob. We thank Allyn Tennant for maintaining Web-QDP and HEASARC at NASA GSFC for maintaining Webspec.  We acknowledge support from NASA grant NNX11AE05G (to ZH).


\label{lastpage}

\end{document}